\documentclass{article}[a4paper]
\usepackage{array, tabularx, caption, boldline}
\usepackage{graphicx}
\usepackage{soul}
\usepackage{makecell}
\usepackage[utf8]{inputenc}
\usepackage{lmodern,textcomp}
\usepackage{hyperref}
\hypersetup{
    colorlinks=true,
    linkcolor=blue,
    filecolor=magenta,      
    urlcolor=blue,
    citecolor=blue
}

\usepackage{float}
\usepackage{cite}
\usepackage{ifthen, mciteplus} 
\usepackage{geometry}
\usepackage{booktabs}
\usepackage{multirow}
\usepackage{amssymb}
\usepackage{amsmath}
\usepackage{comment}
\usepackage{caption}
\usepackage{subcaption}
\usepackage{tabularx}
\usepackage[bottom]{footmisc}

\usepackage{url}
\usepackage{placeins}
\newboolean{pdflatex}
\setboolean{pdflatex}{true} 

\newboolean{articletitles}
\setboolean{articletitles}{true} 
\newboolean{uprightparticles}
\setboolean{uprightparticles}{false} 
\newboolean{inbibliography}
\setboolean{inbibliography}{true} 

\usepackage{microtype}
\usepackage{lineno}  
\usepackage{xspace} 
\usepackage{caption} 

\title{Prospects of searches for invisible $B$-meson decays at FCC-ee}
\author{P.~Alvarez Cartelle$^{1}$, M.~Kenzie$^{1}$, R.~Mangrulkar$^{1}$, A.~R.~Wiederhold$^{2}$, E.~Wood$^{1\dagger}$}
\date{}

\usepackage{xspace} 
\usepackage{upgreek}
\usepackage{amsmath}
\usepackage{color}

\def\binv{\ensuremath{\BdorBs\to \neu\neub}\xspace}
\def\bdinv{\ensuremath{\Bd\to \neu\neub}\xspace}
\def\bsinv{\ensuremath{\Bs\to \neu\neub}\xspace}
\def\binvinv{\ensuremath{\BdorBs \to \neu\neub\neu\neub}\xspace}

\def\invisible{\ensuremath{{\rm{invisible}}}\xspace}
\def\binvisible{\ensuremath{{\BdorBs \to \invisible}}\xspace}
\def\binvisibleg{\ensuremath{{\BdorBs \to \invisible+\g}}\xspace}
\def\binvisiblepg{\ensuremath{{\BdorBs \to \invisible~(+\g)}}\xspace}
\def\bdinvisible{\ensuremath{\Bd \to \invisible}\xspace}
\def\bsinvisible{\ensuremath{\Bs \to \invisible}\xspace}

\def\Zbb{\ensuremath{\Z\to\bquark\bquarkbar}\xspace}
\def\Zcc{\ensuremath{\Z\to\cquark\cquarkbar}\xspace}
\def\Zss{\ensuremath{\Z\to\squark\squarkbar}\xspace}
\def\Zdd{\ensuremath{\Z\to\dquark\dquarkbar}\xspace}
\def\Zuu{\ensuremath{\Z\to\uquark\uquarkbar}\xspace}
\def\Zud{\ensuremath{\Z\to\dquark\dquarkbar+\Z\to\uquark\uquarkbar}\xspace}
\def\Zqq{\ensuremath{\Z\to\quark\quarkbar}\xspace}
\def\Ztautau{\ensuremath{\Z\to\taup\taum}\xspace}
\def\Zmumu{\ensuremath{\Z\to\mup\mun}\xspace}
\def\Zee{\ensuremath{\Z\to\ep\en}\xspace}

\def\dd{\ensuremath{\dquark\dquarkbar}\xspace}
\def\uu{\ensuremath{\uquark\uquarkbar}\xspace}

\def\ud{\ensuremath{\dd+\uu}\xspace}

\def\delphes {\mbox{\textsc{DELPHES}}\xspace}

\def\edmhep{\mbox{\textsc{EDM4hep}}\xspace}

\def\belletwo {\mbox{Belle~II}\xspace}

\def\MagUp {\mbox{\em Mag\kern -0.05em Up}\xspace}

\ifthenelse{\boolean{uprightparticles}}%
{
 
 \def\Pgamma      {\ensuremath{\upgamma}\xspace}

 \def\Pmu         {\ensuremath{\upmu}\xspace}                 
 \def\Pnu         {\ensuremath{\upnu}\xspace}                 
                  
 \def\Ppi         {\ensuremath{\uppi}\xspace}

 \def\Ptau        {\ensuremath{\uptau}\xspace}

 \def\PDelta      {\ensuremath{\Delta}\xspace}                 
 \def\PXi         {\ensuremath{\Xi}\xspace}                 
 \def\PLambda     {\ensuremath{\Lambda}\xspace}                 
 \def\PSigma      {\ensuremath{\Sigma}\xspace}                 
 \def\POmega      {\ensuremath{\Omega}\xspace}                 
 \def\PUpsilon    {\ensuremath{\Upsilon}\xspace}

 \def\PB      {\ensuremath{\mathrm{B}}\xspace}                 
                  
 \def\PD      {\ensuremath{\mathrm{D}}\xspace}

 \def\PK      {\ensuremath{\mathrm{K}}\xspace}

 \def\PZ      {\ensuremath{\mathrm{Z}}\xspace}                 
                  
 \def\Pb      {\ensuremath{\mathrm{b}}\xspace}                 
 \def\Pc      {\ensuremath{\mathrm{c}}\xspace}                 
 \def\Pd      {\ensuremath{\mathrm{d}}\xspace}                 
 \def\Pe      {\ensuremath{\mathrm{e}}\xspace}

 \def\Pi      {\ensuremath{\mathrm{i}}\xspace}

 \def\Pp      {\ensuremath{\mathrm{p}}\xspace}                 
 \def\Pq      {\ensuremath{\mathrm{q}}\xspace}                 
                  
 \def\Ps      {\ensuremath{\mathrm{s}}\xspace}                 
 \def\Pt      {\ensuremath{\mathrm{t}}\xspace}                 
 \def\Pu      {\ensuremath{\mathrm{u}}\xspace}

 \def\thebaroffset{0.0em}
}
{
 
 \def\Pgamma      {\ensuremath{\gamma}\xspace}

 \def\Pmu         {\ensuremath{\mu}\xspace}                 
 \def\Pnu         {\ensuremath{\nu}\xspace}                 
                  
 \def\Ppi         {\ensuremath{\pi}\xspace}

 \def\Ptau        {\ensuremath{\tau}\xspace}

 \mathchardef\PDelta="7101
 \mathchardef\PXi="7104
 \mathchardef\PLambda="7103
 \mathchardef\PSigma="7106
 \mathchardef\POmega="710A
 \mathchardef\PUpsilon="7107
                  
 \def\PB      {\ensuremath{B}\xspace}                 
                  
 \def\PD      {\ensuremath{D}\xspace}

 \def\PK      {\ensuremath{K}\xspace}

 \def\PZ      {\ensuremath{Z}\xspace}                 
                  
 \def\Pb      {\ensuremath{b}\xspace}                 
 \def\Pc      {\ensuremath{c}\xspace}                 
 \def\Pd      {\ensuremath{d}\xspace}                 
 \def\Pe      {\ensuremath{e}\xspace}

 \def\Pi      {\ensuremath{i}\xspace}

 \def\Pp      {\ensuremath{p}\xspace}                 
 \def\Pq      {\ensuremath{q}\xspace}                 
                  
 \def\Ps      {\ensuremath{s}\xspace}                 
 \def\Pt      {\ensuremath{t}\xspace}                 
 \def\Pu      {\ensuremath{u}\xspace}

 \def\thebaroffset{0.18em}
}
\newcommand{\offsetoverline}[2][\thebaroffset]{\kern #1\overline{\kern -#1 #2}}%

\makeatletter
\ifcase \@ptsize \relax
  \newcommand{\miniscule}{\@setfontsize\miniscule{4}{5}}
\or
  \newcommand{\miniscule}{\@setfontsize\miniscule{5}{6}}
\or
  \newcommand{\miniscule}{\@setfontsize\miniscule{5}{6}}
\fi
\makeatother

\DeclareRobustCommand{\optbar}[1]{\shortstack{{\miniscule (\rule[.5ex]{1.25em}{.18mm})}
  \\ [-.7ex] $#1$}}

\def\electron   {{\ensuremath{\Pe}}\xspace}
\def\en         {{\ensuremath{\Pe^-}}\xspace}   
\def\ep         {{\ensuremath{\Pe^+}}\xspace}

\def\muon       {{\ensuremath{\Pmu}}\xspace}
\def\mup        {{\ensuremath{\Pmu^+}}\xspace}
\def\mun        {{\ensuremath{\Pmu^-}}\xspace} 

\def\taup       {{\ensuremath{\Ptau^+}}\xspace}
\def\taum       {{\ensuremath{\Ptau^-}}\xspace}

\def\ellm       {{\ensuremath{\ell^-}}\xspace}
\def\ellp       {{\ensuremath{\ell^+}}\xspace}

\def\neu        {{\ensuremath{\Pnu}}\xspace}
\def\neub       {{\ensuremath{\overline{\Pnu}}}\xspace}
\def\neue       {{\ensuremath{\neu_e}}\xspace}

\def\neum       {{\ensuremath{\neu_\mu}}\xspace}

\def\neut       {{\ensuremath{\neu_\tau}}\xspace}
\def\neutb      {{\ensuremath{\neub_\tau}}\xspace}
\def\neul       {{\ensuremath{\neu_\ell}}\xspace}

\def\g      {{\ensuremath{\Pgamma}}\xspace}

\def\Z      {{\ensuremath{\PZ}}\xspace}

\def\quark     {{\ensuremath{\Pq}}\xspace}
\def\quarkbar  {{\ensuremath{\overline \quark}}\xspace}

\def\uquark    {{\ensuremath{\Pu}}\xspace}
\def\uquarkbar {{\ensuremath{\overline \uquark}}\xspace}

\def\dquark    {{\ensuremath{\Pd}}\xspace}
\def\dquarkbar {{\ensuremath{\overline \dquark}}\xspace}

\def\squark    {{\ensuremath{\Ps}}\xspace}
\def\squarkbar {{\ensuremath{\overline \squark}}\xspace}
\def\ssbar     {{\ensuremath{\squark\squarkbar}}\xspace}
\def\cquark    {{\ensuremath{\Pc}}\xspace}
\def\cquarkbar {{\ensuremath{\overline \cquark}}\xspace}
\def\ccbar     {{\ensuremath{\cquark\cquarkbar}}\xspace}
\def\bquark    {{\ensuremath{\Pb}}\xspace}
\def\bquarkbar {{\ensuremath{\overline \bquark}}\xspace}
\def\bbbar     {{\ensuremath{\bquark\bquarkbar}}\xspace}
\def\tquark    {{\ensuremath{\Pt}}\xspace}
\def\tquarkbar {{\ensuremath{\overline \tquark}}\xspace}

\def\pion   {{\ensuremath{\Ppi}}\xspace}
\def\piz    {{\ensuremath{\pion^0}}\xspace}
\def\pip    {{\ensuremath{\pion^+}}\xspace}
\def\pim    {{\ensuremath{\pion^-}}\xspace}

\def\hp     {{\ensuremath{h^+}}\xspace}

\def\kaon    {{\ensuremath{\PK}}\xspace}
\def\Kbar    {{\ensuremath{\offsetoverline{\PK}}}\xspace}

\def\KorKbar {\kern \thebaroffset\optbar{\kern -\thebaroffset \PK}{}\xspace}
\def\Kz      {{\ensuremath{\kaon^0}}\xspace}
\def\Kzb     {{\ensuremath{\Kbar{}^0}}\xspace}
\def\Kp      {{\ensuremath{\kaon^+}}\xspace}

\def\Kpm     {{\ensuremath{\kaon^\pm}}\xspace}

\def\KS      {{\ensuremath{\kaon^0_{\mathrm{S}}}}\xspace}

\def\D       {{\ensuremath{\PD}}\xspace}

\def\DorDbar {\kern \thebaroffset\optbar{\kern -\thebaroffset \PD}\xspace}
\def\Dz      {{\ensuremath{\D^0}}\xspace}

\def\Dp      {{\ensuremath{\D^+}}\xspace}

\def\Dsp     {{\ensuremath{\D^+_\squark}}\xspace}

\def\B       {{\ensuremath{\PB}}\xspace}

\def\BorBbar {\kern \thebaroffset\optbar{\kern -\thebaroffset \PB}\xspace}

\def\Bd      {{\ensuremath{\B^0}}\xspace}

\def\BdorBdbar {\kern \thebaroffset\optbar{\kern -\thebaroffset \Bd}\xspace}
\def\Bu      {{\ensuremath{\B^+}}\xspace}

\def\Bp      {{\ensuremath{\Bu}}\xspace}

\def\Bs      {{\ensuremath{\B^0_\squark}}\xspace}

\def\BsorBsbar {\kern \thebaroffset\optbar{\kern -\thebaroffset \Bs}\xspace}

\def\BdorBs  {{\ensuremath{\B{}_{(\squark)}^0}}\xspace}

\def\BuorBc  {{\ensuremath{\B{}_{(\cquark)}^+}}\xspace}

\def\Y#1S{\ensuremath{\PUpsilon{(#1S)}}\xspace}

\def\proton      {{\ensuremath{\Pp}}\xspace}

\def\porpbar     {\kern \thebaroffset\optbar{\kern -\thebaroffset \proton}\xspace}
\def\Lz          {{\ensuremath{\PLambda}}\xspace}

\def\LorLbar     {\kern \thebaroffset\optbar{\kern -\thebaroffset \PLambda}\xspace}

\def\Lb           {{\ensuremath{\Lz^0_\bquark}}\xspace}

\def\BF         {{\ensuremath{\mathcal{B}}}\xspace}

\def\to                 {\ensuremath{\rightarrow}\xspace}

\def\AT#1     {\ensuremath{A_{\mathrm{T}}^{#1}}\xspace}           

\def\C#1      {\ensuremath{\mathcal{C}_{#1}}\xspace}                       
\def\Cp#1     {\ensuremath{\mathcal{C}_{#1}^{'}}\xspace}                    
\def\Ceff#1   {\ensuremath{\mathcal{C}_{#1}^{\mathrm{(eff)}}}\xspace}        
\def\Cpeff#1  {\ensuremath{\mathcal{C}_{#1}^{'\mathrm{(eff)}}}\xspace}       
\def\Ope#1    {\ensuremath{\mathcal{O}_{#1}}\xspace}                       
\def\Opep#1   {\ensuremath{\mathcal{O}_{#1}^{'}}\xspace}                    

\newcommand{\nospaceunit}[1]{\ensuremath{\text{#1}}}       
\newcommand{\aunit}[1]{\ensuremath{\text{\,#1}}}       

\newcommand{\tev}{\aunit{Te\kern -0.1em V}\xspace}
\newcommand{\gev}{\aunit{Ge\kern -0.1em V}\xspace}
\newcommand{\mev}{\aunit{Me\kern -0.1em V}\xspace}
\newcommand{\kev}{\aunit{ke\kern -0.1em V}\xspace}
\newcommand{\ev}{\aunit{e\kern -0.1em V}\xspace}
\newcommand{\mevc}{\ensuremath{\aunit{Me\kern -0.1em V\!/}c}\xspace}
\newcommand{\gevc}{\ensuremath{\aunit{Ge\kern -0.1em V\!/}c}\xspace}
\newcommand{\mevcc}{\ensuremath{\aunit{Me\kern -0.1em V\!/}c^2}\xspace}
\newcommand{\gevcc}{\ensuremath{\aunit{Ge\kern -0.1em V\!/}c^2}\xspace}

\def\mum  {\ensuremath{\,\upmu\nospaceunit{m}}\xspace}

\def\ab   {\ensuremath{\aunit{ab}}\xspace}
\def\invab   {\ensuremath{\ab^{-1}}\xspace}

\def\gsim{{~\raise.15em\hbox{$>$}\kern-.85em
          \lower.35em\hbox{$\sim$}~}\xspace}
\def\lsim{{~\raise.15em\hbox{$<$}\kern-.85em
          \lower.35em\hbox{$\sim$}~}\xspace}

\def\evtgen     {\mbox{\textsc{EvtGen}}\xspace}

\def\photos     {\mbox{\textsc{Photos}}\xspace}

\def\pythia     {\mbox{\textsc{Pythia}}\xspace}

\def\tell1  {TELL1\xspace}
\def\ukl1   {UKL1\xspace}

\newcommand{\ie}{\mbox{\itshape i.e.}\xspace}

\def\BF {{\ensuremath{\mathcal{B}}}\xspace}

\begin{document}

\begin{flushright}
    \href{https://doi.org/10.1007/JHEP04(2026)011}{DOI: 10.1007/JHEP04(2026)011}\\
    \href{https://doi.org/10.17181/q9yx5-9cx10}{DOI: 10.17181/q9yx5-9cx10} 

\end{flushright}
\vspace*{3\baselineskip}

{\let\newpage\relax\maketitle}

\begin{center}{\footnotesize \it
\noindent
$^{1}$Cavendish Laboratory, University of Cambridge, JJ Thomson Avenue, Cambridge, U.K. \\
$^{2}$Department of Physics and Astronomy, University of Manchester, Oxford Road, Manchester, U.K.\\
$^{\dagger}$Corresponding Author
 }

\vspace{0.5cm}

{\footnotesize
{{Email:~}}{\bf\color{blue} paula.alvarez@cern.ch, matthew.kenzie@cern.ch, \\ 
ritwik.mangrulkar@cern.ch,  aidan.richard.wiederhold@cern.ch, ella.wood@cern.ch}
}
\end{center}
\smallskip

\begin{center}
Published in JHEP \textbf{04} (2026) 011
\end{center}
\smallskip

\hrule
\begin{abstract}\noindent
    We investigate the physics reach and potential for the study of $B$-meson decays to invisible final states at the Future Circular Collider running electron-positron collisions at the $Z$ pole (FCC-ee). 
    Signal and background candidates are simulated for a proposed multipurpose detector, including inclusive contributions from $Z$ decays to leptons or quarks. Signal candidates are selected by a mixture of rectangular cuts and a multiclass boosted decision tree classifier. 

    We determine that branching fractions above $7.6\times10^{-9}$ ($9.7\times10^{-9}$) would be excluded at 90\% (95\%) confidence level, and branching fractions above $3.0\times10^{-8}$ would be within discovery reach at FCC-ee if $6\times 10^{12}$ $Z$ bosons are produced. 

\end{abstract}

\vspace*{0.5em}

\hrule

\clearpage 



\section{Introduction}
\label{sec:introduction}

The study of rare and invisible decays of $B$ mesons offers a sensitive probe for physics beyond the Standard Model (SM).
In particular, searches for invisible decays, such as \binv, are of great interest given their heavy suppression in the SM.
Extensions to the SM, for example those including dark matter~\cite{Bird:2004ts}, light neutralino~\cite{OLeary:2009cqb} or axion-like~\cite{Alonso-Alvarez:2023mgc} candidates, can significantly enhance the rates of these processes.

\begin{figure}[b]
    \centering
    \includegraphics[width=0.49\linewidth]{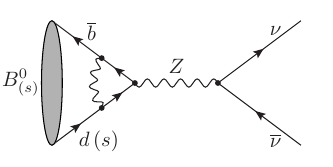}
    \includegraphics[width=0.49\linewidth]{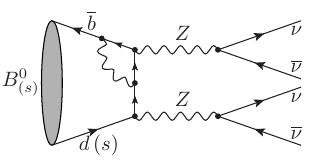}
    \caption{Example Feynman diagrams for the \binv (left) and \binvinv (right) decays. The $W$-boson loop in the \binvinv decay can appear on either initial state quark.}
    \label{fig:feynmans}
\end{figure}

Decays of neutral $B$ mesons (either \Bd or \Bs) into neutrino pairs, \binv, are heavily loop and helicity suppressed in the SM.\footnote{Charge conjugation is implied throughout, unless otherwise explicitly stated.}
The helicity suppression factor is proportional to $(m_\neu / m_\B)^2$ resulting in expected branching fractions of $\mathcal{O}(10^{-25})$~\cite{Bhattacharya:2018msv, Lu:1996et, Bortolato:2020bgy, Badin:2010uh}. 
While experimentally identical, decays of \binvinv do not suffer from the same level of suppression and thus are predicted to have higher branching fractions of $\mathcal{O}(10^{-15})$~\cite{Bhattacharya:2018msv}. Some example Feynman diagrams for these SM processes are shown in Fig.~\ref{fig:feynmans}. For \binvisibleg, the branching fraction is further enhanced to $\mathcal{O}(10^{-9})$~\cite{Lu:1996et}. 
A combined \binvisiblepg analysis could therefore be envisaged, similar to that performed by \textsc{BaBar}~\cite{BaBar:2012yut}, where single photons detected above a given energy threshold are used to classify events as \binvisible or \binvisibleg. In the \textsc{BaBar} analysis, the effect of cross-feed between the two channels was found to be negligible; however further investigation into this effect lies beyond the scope of this work. 
Regardless, the predicted branching fractions for invisible \BdorBs decays in the SM are so small that searches for them effectively provide a null test for the SM. A significant signal from any conceivable facility in the near future would provide striking evidence for new physics (NP).

From a theory perspective \binv decays are exceptionally clean. 
As purely leptonic annihilation processes, they are free from hadronic form factors, and their decay rates depend only  on the decay constants, $f_{\BdorBs}$. As a result, these modes are largely unaffected by significant final-state QED corrections, long-distance QCD contributions and charm-loop effects, which often complicate theoretical predictions in other rare $B$-meson decays.
These modes are therefore sensitive probes of physics beyond the SM, as they test couplings to all three neutrino generations in a theoretically well-controlled environment.

Recently, the \belletwo collaboration has reported a measurement of the related transition decay $\Bp\to\Kp \neu\neub$~\cite{Belle-II:2023esi}, showing a central value above the SM expectation, albeit with limited statistical significance. 
If this deviation persists with additional data, it would point to NP affecting $\bquark\to\squark$ transitions with invisible final states.
In such a scenario, complementary measurements and limits on fully invisible $B$ decays provide valuable additional information, as they probe different operator structures and flavour topologies. 
In particular, while transition decays constrain flavour-changing currents, invisible \BdorBs decays are sensitive to annihilation processes that are absent or highly suppressed in the SM.

Moreover, enhancements in invisible \BdorBs decays need not necessarily be accompanied by observable effects in $\bquark\to\squark\neu\neub$ transitions. 
This situation can arise in models where the invisible final state consists of non-SM particles, such as dark-sector states or long-lived neutral particles, rather than SM neutrinos.
Examples include scenarios mediated by leptophobic vector portals ($Z$ portals)~\cite{Bird:2004ts, Leroux:2001fx, Baek:2006bv}, Higgs-portal scalars~\cite{Patt:2006fw}, or axion-like particles~\cite{Bauer:2022rwf,Alonso-Alvarez:2023mgc}, in which neutral $B$ mesons can decay invisibly without inducing sizeable couplings to neutrinos.
Searches for invisible $B$ decays therefore probe a broad and well-motivated class of NP scenarios that are complementary to existing, and future, measurements of rare semileptonic $B$ decays.

The current best limits on the branching fraction of \bdinvisible decays comes from \textsc{BaBar} which finds $\BF(\bdinvisible) < 2.4\times 10^{-5}$ and $\BF( \bdinvisible +\gamma) < 1.7 \times 10^{-5}$ at 90\%~CL~\cite{BaBar:2012yut}. 
More recently, a reinterpretation of the original ALEPH data finds $\BF(\bdinvisible) < 1.4\times 10^{-4}$ and $\BF(\bsinvisible) < 5.6\times 10^{-4}$, both at 90\%~CL~\cite{Alonso-Alvarez:2023mgc}.
Looking to the future, with the full $50\invab$ $\Upsilon(4S)$ ($5\invab$ $\Upsilon(5S)$) dataset, \belletwo expect to reach sensitivities down to $1.5 \times 10^{-6}$ ($1.1 \times 10^{-5}$) for the \bdinvisible (\bsinvisible) branching fraction at 90\%~CL~\cite{Belle-II:2018jsg}.
Due to the final state being entirely invisible, measurements of these processes are impossible at LHCb.
Consequently the only potential for improvement on the \belletwo results is from future facilities.

This paper describes a feasibility study for the prospects of searching for invisible $B$-meson decays at the Future Circular Collider (FCC) running in electron-positron mode at the $Z$ threshold (FCC-ee).

\section{Experimental environment}
\label{sec:experimental-setup}

For our experimental analysis, we follow much of the procedure developed and outlined in Refs.~\cite{Amhis:2021cfy,Amhis:2023mpj,Zuo:2023dzn}.
Here we give a brief description of the collider and detector environment that has been assumed for this study.

\subsection{FCC-ee}

The FCC~\cite{FCC:2018byv} is a proposal for the next generation state-of-the-art particle research facility.
Following the FCC feasibility study~\cite{FCC:2025lpp,FCC:2025uan,FCC:2025jtd} there are ongoing efforts to investigate the full physics potential of such a machine, which would be built in a new 90.7~km tunnel near CERN, with capabilities of running in successive stages of $e^+e^-$ or $pp$ mode.
The \ep\en machine, FCC-ee~\cite{FCC:2018evy}, would run at a range of centre-of-mass energies, $\sqrt{s}$, between 91\gev (\ie the $Z$ pole) and 365\gev (\ie the \tquark\tquarkbar threshold).
FCC-ee offers unprecedented opportunities to study every known particle of the SM in exquisite detail. 
Beyond its capabilities as an electroweak precision machine, there is scope for world's-best measurements in the beauty (\bquark-quark), charm (\cquark-quark) and tau ($\tau$-lepton) sectors with the vast statistics anticipated to be taken at the $Z$ pole. 
This so called ``Tera-$Z$" run would produce $\mathcal{O}(10^{12})$ $Z$ bosons per experiment, which have a high branching fraction to both \bquark\bquarkbar (0.15) and \cquark\cquarkbar (0.12) pairs~\cite{PDG2022}.
In contrast with other proposed future colliders, the low-energy operation of FCC-ee allows for data samples at the $Z$ pole orders of magnitude larger than those achievable elsewhere, thereby allowing for substantially more precise measurements~\cite{Bambade:2019fyw}.
Another advantage of a circular, as opposed to linear,  collider layout is that collisions can be delivered to multiple interaction regions simultaneously, which allows for a variety of different detector design choices.

Our study assumes that $6 \times 10^{12}$ $Z$ bosons will be produced at FCC-ee, integrated across four experiments.
This would provide a sample of approximately 720B \Bd mesons and 180B \Bs mesons in a clean experimental environment~\cite{Monteil:2021ith}.

\subsection{Detector response}

Monte Carlo (MC) event samples are used to simulate the response of the detector to various different physics processes. 
The procedure for event generation and simulation of the detector response is identical to that described in Ref.~\cite{Amhis:2021cfy}.
In summary, events are generated under nominal FCC-ee conditions using \pythia~\cite{Sjostrand:2014zea}, with unstable particles decayed using \evtgen~\cite{Lange:2001uf} and final-state radiation generated by \photos~\cite{Davidson:2010ew}.
The detector configuration under consideration is the Innovative Detector for Electron-positron Accelerators (IDEA) concept~\cite{IDEAStudyGroup:2025gbt}.
It consists of a silicon pixel vertex detector, a large-volume, extremely light, short-drift wire chamber surrounded by a layer of silicon microstrip detectors, a thin low-mass superconducting solenoid coil, a preshower, a dual-readout calorimeter, and muon chambers within the magnet return yoke~\cite{FCC:2018evy}.
The detector response is simulated using the \delphes package~\cite{deFavereau:2013fsa} with the configuration card in Ref.~\cite{giovanni_marchiori_2025_15544272} interfaced to the common \edmhep data format~\cite{valentin_volkl_2021_4785063}.

\subsection{Simulation samples}\label{sec:MCsamples}

Our study employs various different MC simulation samples to mimic the expected signal and background distributions at FCC-ee. 
We make use of inclusive samples of \Zbb, \Zcc, \Zss, \Zqq (where $q$ is one of the light quarks, $q\in\{\uquark, \dquark\}$), \Ztautau, \Zmumu and \Zee as proxies for the total expected background.
Almost all of the backgrounds from $Z$ decays to charged leptons can be entirely removed with relatively trivial selection requirements (detailed further in Sec.~\ref{sec:preselection}). In addition, due to the low statistics remaining in these inclusive samples after the full selection, a cocktail of exclusive \BuorBc\to~\ellp\neul MC samples is studied in further detail as outlined in Sec.~\ref{sec:b2lnu}.
We then make use of dedicated exclusive \bdinv and \bsinv samples for \binvisible decays.
The simulated samples contain an admixture of both \bquark-hadron flavours \ie charge conjugation is implied. 

\subsection{Analysis framework and implementation}\label{sec:framework}

We exploit a similar analysis framework to that used in Refs.~\cite{Amhis:2021cfy,Amhis:2023mpj,Zuo:2023dzn}.
The fit for the primary vertex (PV) is performed using only reconstructed tracks and does not rely on any underlying MC truth information.
For any displaced vertices (DVs) the vertex fits are initially seeded using MC truth information. 
For each MC vertex, the MC tracks produced from it are matched to their corresponding reconstructed tracks upon which a vertex fit is  performed.
Whilst this accounts for the impact of imperfect resolution on tracks to the resulting vertex position, it assumes that each reconstructed track is correctly associated to its origin vertex.

In addition, perfect particle identification (PID) is also assumed throughout, with each reconstructed charged track assigned a mass based on the underlying MC truth.
This is expected to have a minimal impact on the performance of the analysis because we do not rely heavily on correct PID. 
However, one variable used to reject background is the reconstructed mass of certain vertices, including the PV.
In reality the resolution of this variable will have some dependence on the PID of the tracks from the vertex.
In principle, the overall sensitivity of this analysis could be improved by additionally exploiting PID information, which is discussed in more detail in Sec.~\ref{sec:separating_bd_and_bs}.

\section{Analysis}
\label{sec:analysis}

In order to obtain an estimate for the expected sensitivity to invisible $B$-meson decays, we optimise a two-stage selection procedure based on an initial preselection followed by a multiclass boosted decision tree (BDT) classifier.
The preselection removes the majority of background from $Z\to\ellp\ellm$, for $\ell \in \{\electron, \muon, \tau\}$, and a considerable fraction of the hadronic background.
The BDT is trained to distinguish between the signal candidates of interest and the inclusive backgrounds from \Zbb, \Zcc, \Zss and \Zqq, for $q\in\{\uquark,\dquark\}$, which are categorised into \emph{heavy} (\Zbb and \Zcc) and \emph{light} (\Zss, \Zdd, \Zuu).

\begin{figure}
     \centering
     \begin{subfigure}[t]{0.325\textwidth}
         \centering
         \caption{\bdinv}
         \includegraphics[width=\textwidth]{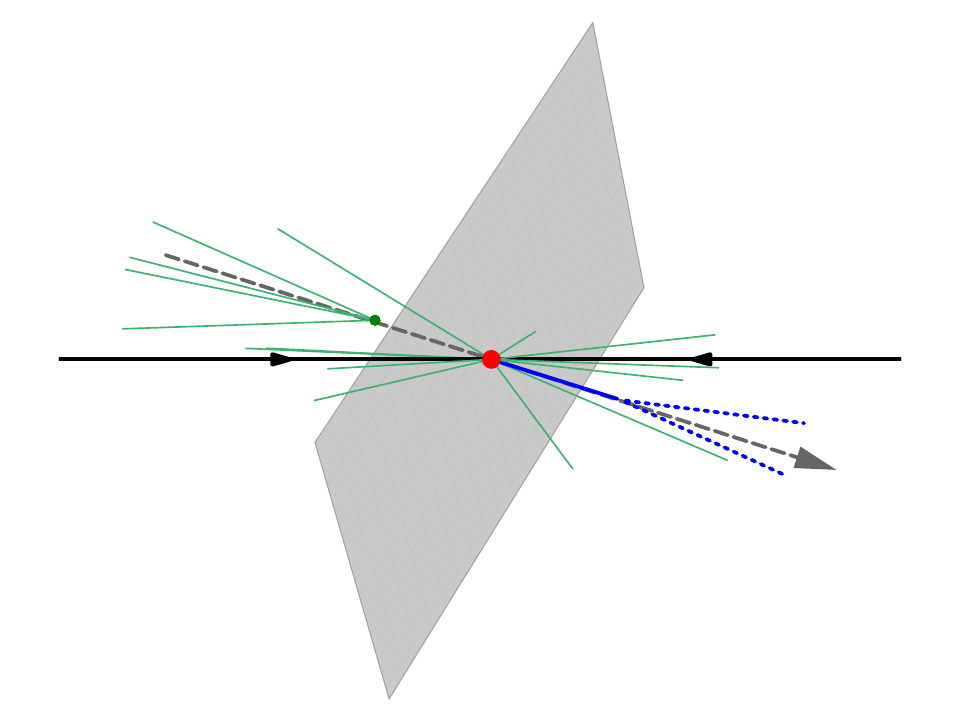}
     \end{subfigure}
     \hfill
     \begin{subfigure}[t]{0.325\textwidth}
         \centering
          \caption{\Zbb}
         \includegraphics[width=\textwidth]{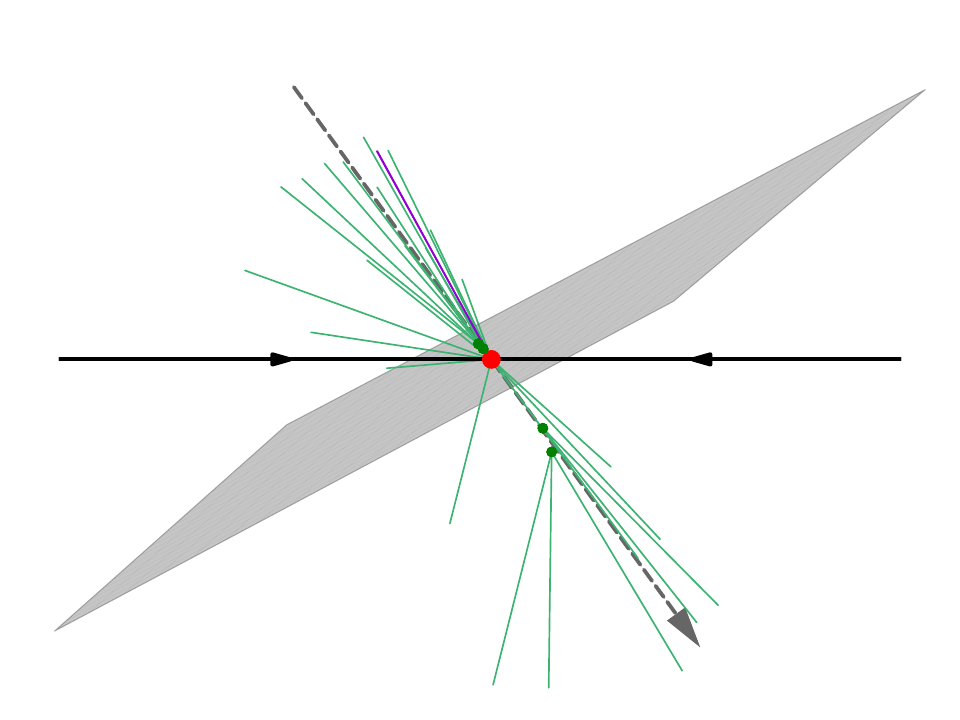}
     \end{subfigure}
     \hfill
     \begin{subfigure}[t]{0.325\textwidth}
         \centering
          \caption{\Zdd}
         \includegraphics[width=\textwidth]{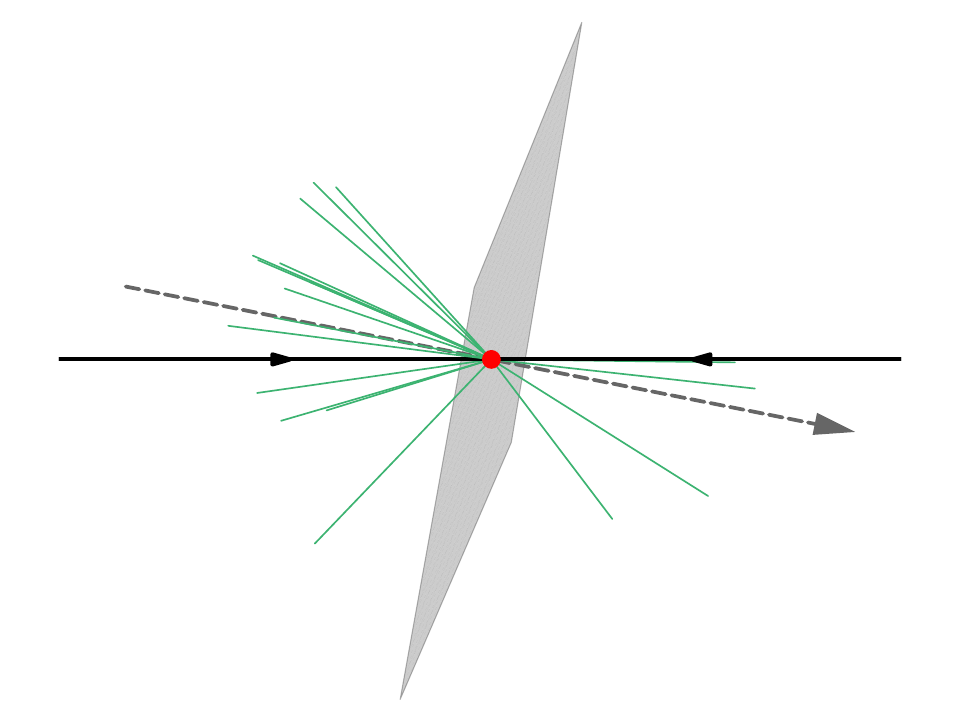}
     \end{subfigure}
     \hfill
     \begin{subfigure}[t]{0.7\textwidth}
         \centering
         \includegraphics[width=\textwidth]{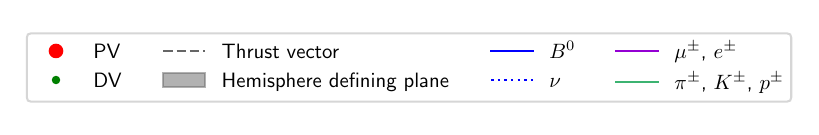}
     \end{subfigure}        \caption{Demonstrative event displays of a typical signal event (left), heavy \Zbb (middle) and light \Zdd (right) hadronic background events passing the preselection cuts.}
         \label{fig:event_display}
\end{figure}

Event displays for typical signal, heavy and light hadronic background decays are shown in Fig.~\ref{fig:event_display}.
The $Z$ bosons at FCC-ee are produced at threshold and are thus approximately at rest.
The subsequent two-body decay into quarks is back-to-back in the $Z$ rest frame and thus almost back-to-back in the lab frame.
One of the key signatures of the signal decays is the presence of large missing energy in the direction of the $B$-meson candidate due to the invisible final state.
Consequently, a typical signal event will have a relatively large imbalance of missing energy between the signal side of the \Zbb event and the non-signal side. 
For a typical \Zbb background event, the missing energy imbalance is significantly smaller. 
In order to determine the imbalance between the signal side and the non-signal side we divide events (on a per-event basis) into two hemispheres, each corresponding to one of the two \bquark quarks produced from the \Z decay.

The hemispheres, pictorially represented in Fig.~\ref{fig:event_display}, are defined using the plane normal to the \emph{thrust axis}, which is defined by the unit vector, $\hat{\mathbf{n}}$, that maximises,
\begin{equation}
    \label{eq:ThrustFOM}
    T = \frac{\sum_i | \mathbf{p}_i \cdot \hat{\mathbf{n}} |}{\sum_i |\mathbf{p}_i| },
\end{equation}
where $\mathbf{p}_i$ is the momentum vector of the $i^\text{th}$ reconstructed particle in the event.
This thrust axis provides a measure of the direction of the quark pair produced from the \Z decay and is defined to point towards the minimum-energy (signal) hemisphere.
Reconstructed particles from each event are then assigned to either hemisphere depending on the angle, $\theta$, between their momentum vector and the thrust axis.
A particle is considered to be in the signal hemisphere (which is expected to have the least total energy) if $\cos(\theta)>0$ and in the non-signal hemisphere if $\cos(\theta)<0$. 
A reconstructed vertex is considered to be in the signal (non-signal) hemisphere if the total reconstructed three-momentum associated with that vertex has $\cos(\theta) > 0$ $(\cos(\theta) < 0)$.
The large boost of the decaying particles typically ensures that this method correctly reconstructs the position of the decay vertex.
An alternative definition for vertex assignment, based on the angle between the thrust axis and the vector joining the PV and DV, has been explored.
This has a 97\% overlap on hemisphere assignment with the former method with the 3\% disagreement primarily driven by the DV resolution and the DV appearing to be behind the PV.
We discard the 3\% of events for which the hemisphere assignment of any vertex does not agree between the two methods.

\begin{figure}
    \centering
    \includegraphics[width=0.49\textwidth]{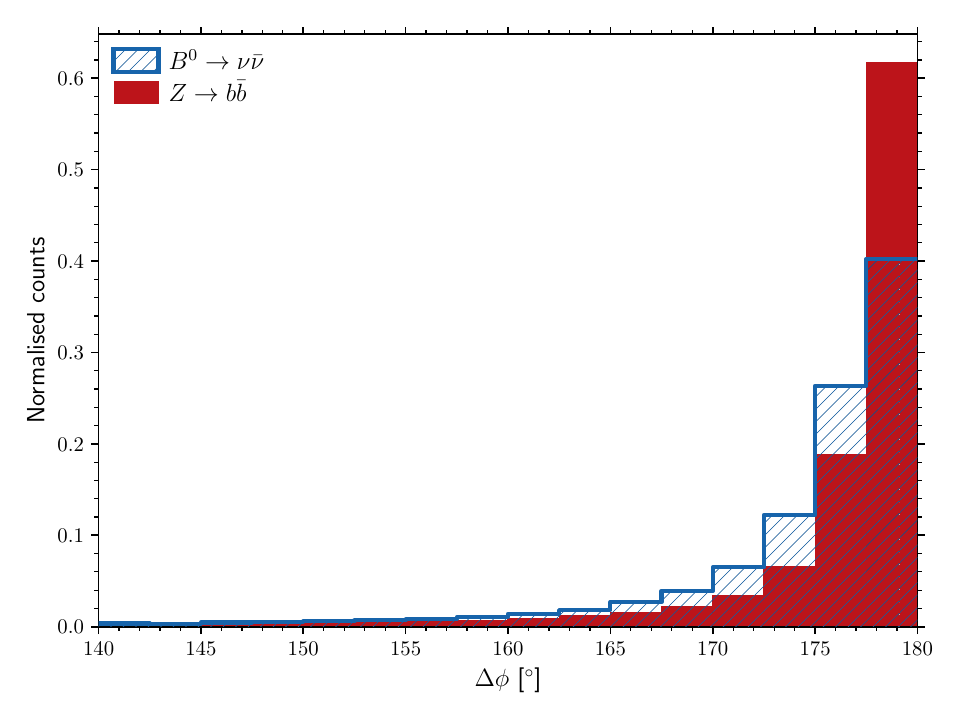}
    \includegraphics[width=0.49\textwidth]{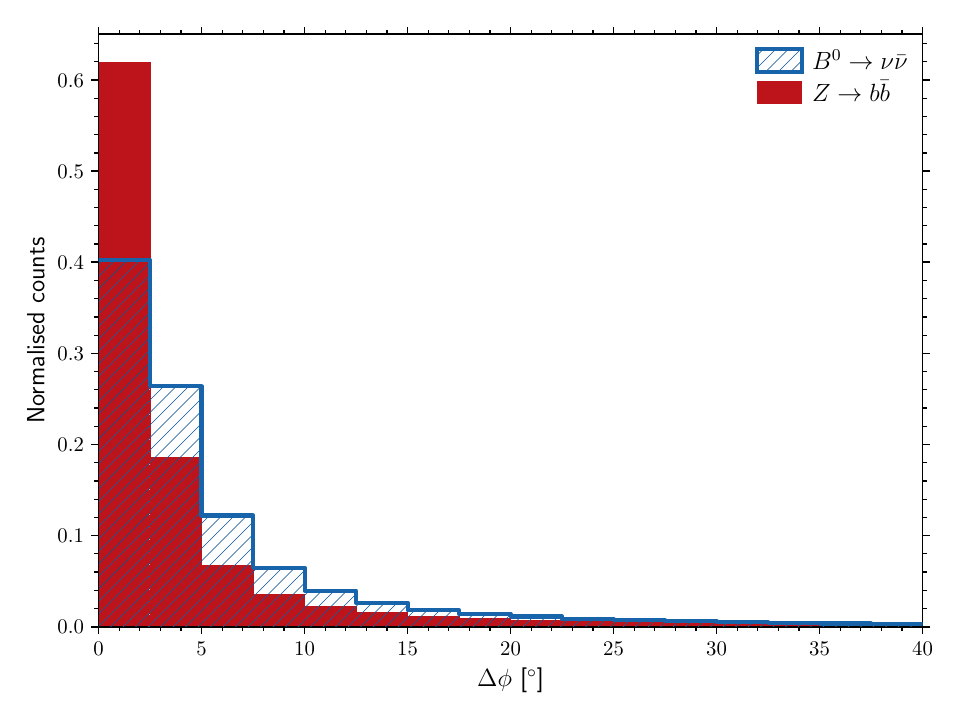}
    \caption{The angle between the reconstructed thrust axis, $\hat{\mathbf{n}}$, and the signal hemisphere \bquark quark (right) and non-signal hemisphere \bquark quark (left) for \bdinv signal (blue hatched) and inclusive \Zbb background (red filled). The distributions are similar for the inclusive backgrounds from \Zcc, \Zss, \Zdd and \Zuu also.}
    \label{fig:thrust_resolution}
\end{figure}

Although there is considerable missing energy in one hemisphere for the signal decays the thrust axis reconstruction resolution is still good and is not biased. 
Figure~\ref{fig:thrust_resolution} shows the angle between the thrust axis vector and the true $\Zbb$ quark flight direction, which suggests a resolution on the thrust axis reconstruction of $\sim 4.9^\circ$ for the \binv signal modes, compared to $\sim 3.1^\circ$ for the hadronic $\Zqq$ background.\footnote{Due to the non-Gaussian nature of the thrust-difference angle distributions, the resolutions are computed using the full width at half the maximum divided by 2.35.}
There is consequently a small chance that a particle is assigned to the wrong hemisphere although as demonstrated further below this method generally works well.

Once the hemispheres are specified by the thrust vector, we define the \emph{signal side} as the hemisphere with the smallest total reconstructed energy.
The other hemisphere we refer to as the \emph{non-signal side} or the \emph{opposite side}.

\subsection{Preselection requirements}
\label{sec:preselection}

Events are required to have at least one reconstructed PV, which demands the PV has at least two charged tracks, and a total reconstructed energy of less than 85\gev, targeting events with significant missing energy.
At least one charged reconstructed particle in the signal hemisphere is required. 
For the signal this targets charged particles produced in association with the signal $B$-meson fragmentation process.
This requirement is still highly efficient for signal but helps to remove background events for which nothing is reconstructed on the signal side and consequently end up looking spuriously like signal events for the multivariate classifier used in the next stage of the selection.
In addition, events with a reconstructed charged lepton, identified in the simulation as either an electron or a muon, in the signal hemisphere are rejected.
This makes the assumption that all electrons and muons are correctly identified. 
This requirement helps to remove a significant fraction of background samples that contain one or more semileptonic decays of beauty or charm hadrons which result in missing energy and an \electron or \muon in the signal hemisphere.
We also require that the reconstructed PV invariant mass is less than 40\gev.
This isolates $Z$ decays to heavy \bquark or \cquark quarks, which typically fly some distance and thus carry mass away from the PV, from $Z$ decays to light \squark, \dquark or \uquark quarks.

Finally, there is a preselection requirement that the particle multiplicity (including reconstructed neutral and charged particles) is more than ten on the non-signal side, which effectively reduces the background from \Ztautau events to negligible levels. 
Decays of $Z$ bosons to two charged leptons are typically low multiplicity on both sides whereas the signal decay will have a relatively high multiplicity on the non-signal side due to the fragmentation process of the other \bquark quark.
The effect of the charged lepton background, in particular from \Ztautau decays, is discussed in more detail in Sec.~\ref{sec:ztautau}.

\begin{figure}
    \centering 
    \includegraphics[width=0.49\textwidth]{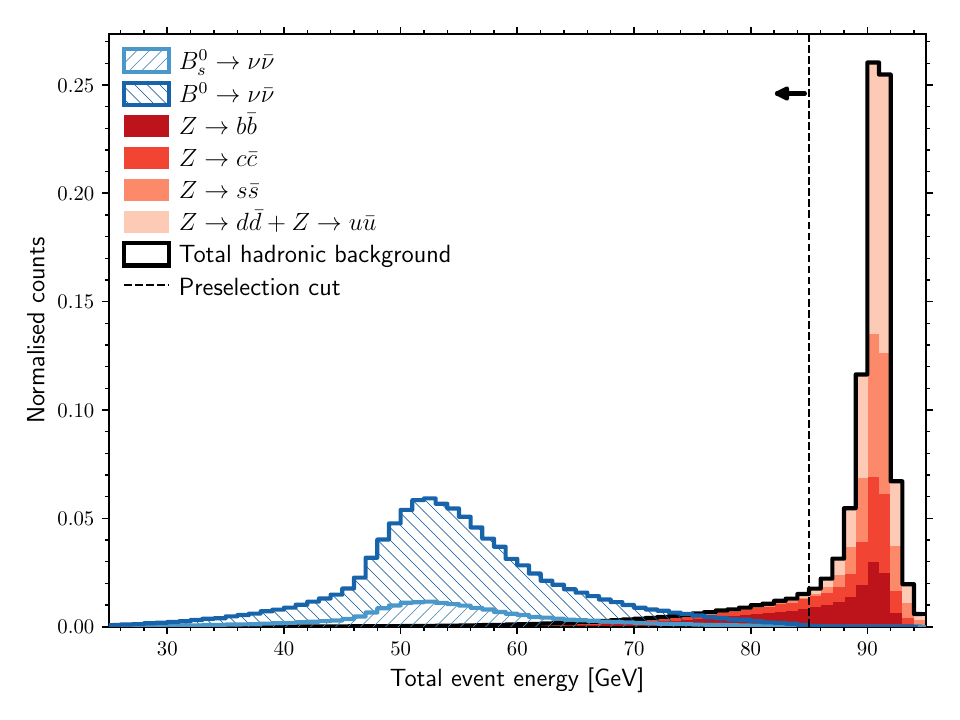}
    \includegraphics[width=0.49\textwidth]{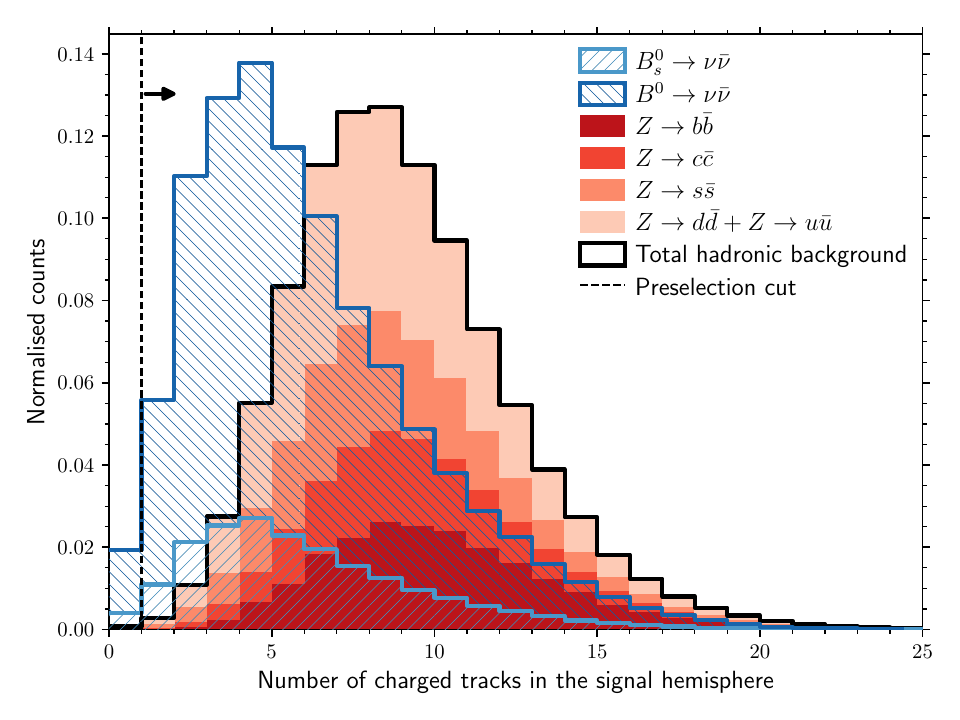} \\
    \includegraphics[width=0.49\textwidth]{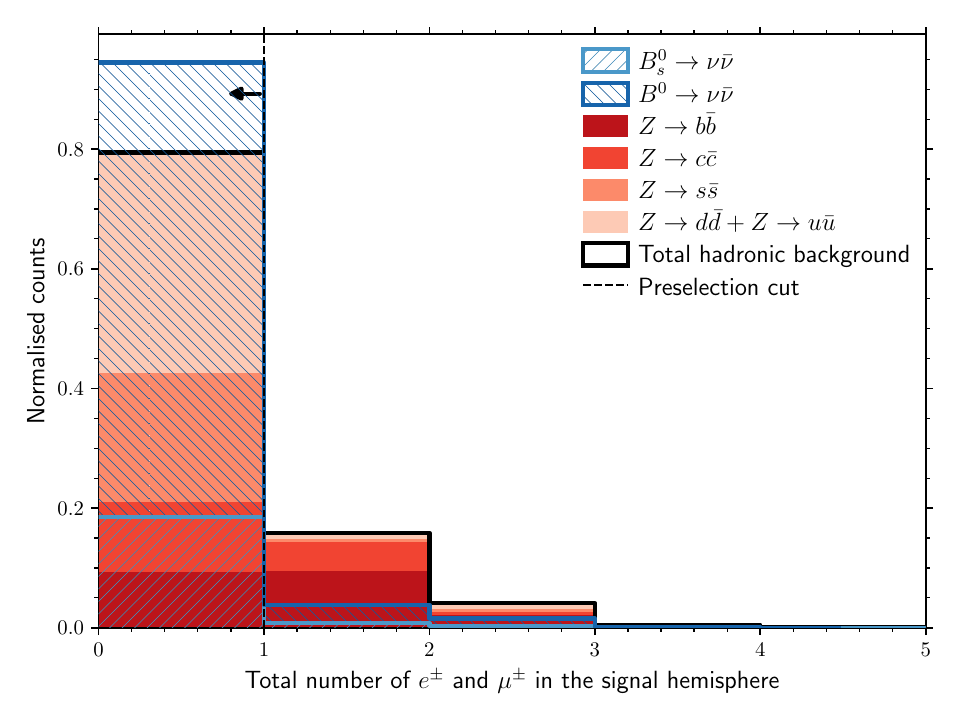}
    \includegraphics[width=0.49\textwidth]{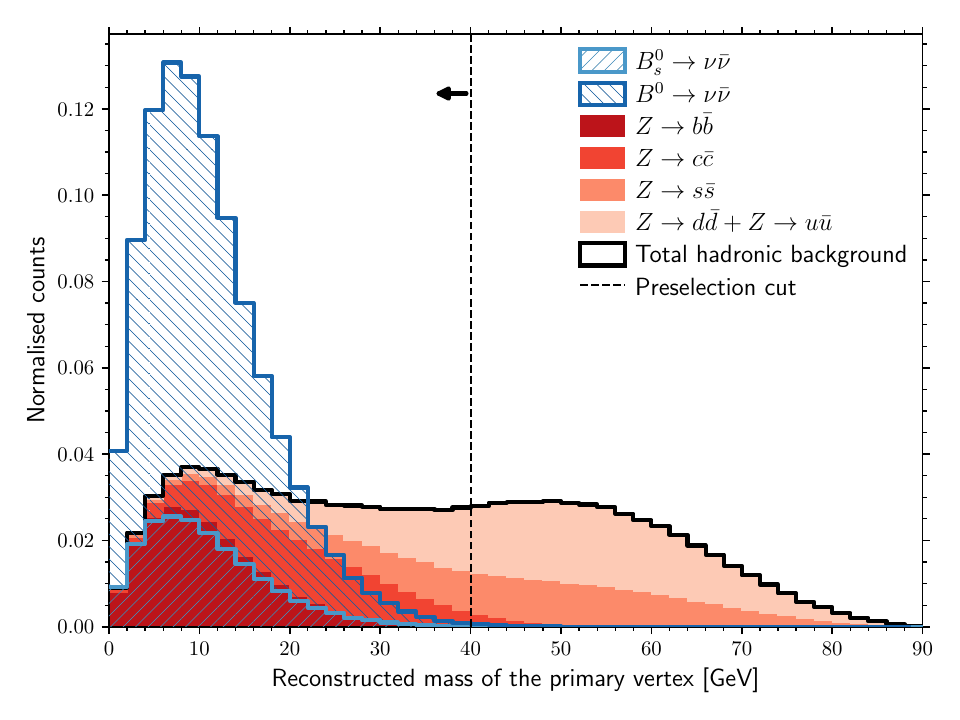}
    \caption{Distributions of the variables used in the preselection requirements including the total reconstructed energy per event (top left), the number of reconstructed charged particles in the signal hemisphere (top right), the number of reconstructed electrons and muons in the signal hemisphere (bottom left) and the reconstructed primary vertex mass (bottom right). 
        The weighted sum of expected \bdinv and \bsinv signal is shown as a stacked histogram in blue with hatched fill. The weighted sum of expected hadronic backgrounds is shown as a stacked histogram in shades of red with the total outlined in black. The preselection requirement is shown by the vertical dashed black line with the small arrow signifying the region which is kept for further analysis.
    }
    \label{fig:preselection_cuts}
\end{figure}

Distributions of the signal and background samples for the preselection variables before any cuts, aside from the requirement of a reconstructed PV, are shown in Fig.~\ref{fig:preselection_cuts}.
Events are appropriately weighted according to the known hadronic \Z branching fractions: 0.1512 (\Zbb), 0.1203 (\Zcc), 0.1584 (\Zss) and 0.2701 (\Zud)~\cite{ParticleDataGroup:2024cfk}, which are expected to be measured with incredibly high precision at FCC-ee~\cite{Rohrig:2025bea}.
The efficiencies of the preselection requirements for each sample are summarised in Table~\ref{tab:preselection_efficiencies}.

\setlength{\tabcolsep}{2pt}
\renewcommand{\arraystretch}{1.2}
\begin{table}
    \centering
    \begin{tabular}{l r c l}
        \textbf{Sample} & \multicolumn{3}{c}{\textbf{Efficiency (\%)}} \\
        \hline
        \bdinv       & $86.592$ & $\pm$ & $0.023$ \\
        \bsinv       & $85.917$ & $\pm$ & $0.025$ \\
        \hline
        \Zbb         & $5.6826$ & $\pm$ & $0.0011$ \\
        \Zcc         & $3.9436$ & $\pm$ & $0.0009$ \\
        \Zss         & $4.1542$ & $\pm$ & $0.0009$ \\
        \Zud$\qquad$ & $2.1885$ & $\pm$ & $0.0007$ \\
        \Ztautau     & ($3.43$ & $\pm$ & $0.06$)$\times 10^{-3}$\\
        \Zmumu       & ($1.2$ & $\pm$ & $0.3$)$\times 10^{-5}$ \\
        \Zee         &($6$ & $\pm$ & $2$)$\times 10^{-6}$ \\

    \end{tabular}
    \caption{Efficiencies of the preselection requirements for each of the samples used in this analysis. The sample labelled as \Zbb includes only background candidates.}
    
    \label{tab:preselection_efficiencies}
\end{table}

\subsection{Multivariate classifier}
\label{sec:bdt}

A multiclass BDT is trained to identify three classes of event which display rather different characteristics in the reconstruction: invisible $B$-meson \emph{signal} decays, \emph{heavy} hadronic backgrounds from \Zbb and \Zcc, and \emph{light} hadronic backgrounds from \Zss, \Zdd and \Zuu.

The signal is characterised by large missing energy on the signal side, along with charged tracks from the $B$-meson hadronisation process and very few displaced tracks or vertices.
This is because the signal \BdorBs meson carries on average 70\% of the \bquark-quark energy~\cite{2011DELPHI} and decays invisibly.
In addition, the signal typically has displaced vertices and many charged tracks on the opposite side which originate from the hadronisation and subsequent decay of the other $b$ quark from the \Z decay.

\begin{figure}
    \centering
    \includegraphics[width=0.49\textwidth]{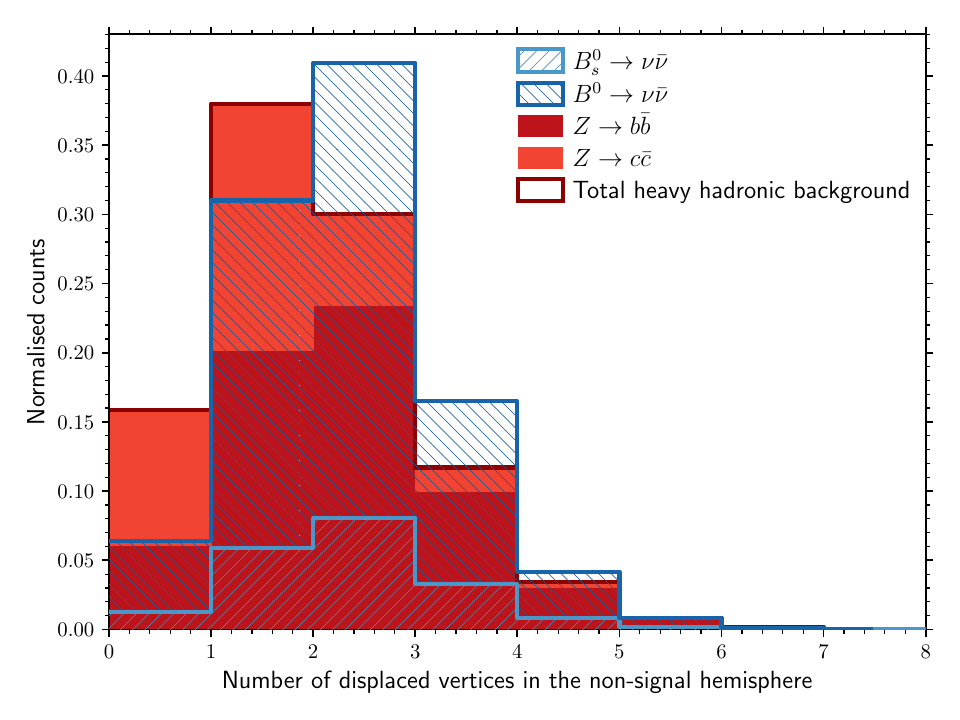}
    \includegraphics[width=0.49\textwidth]{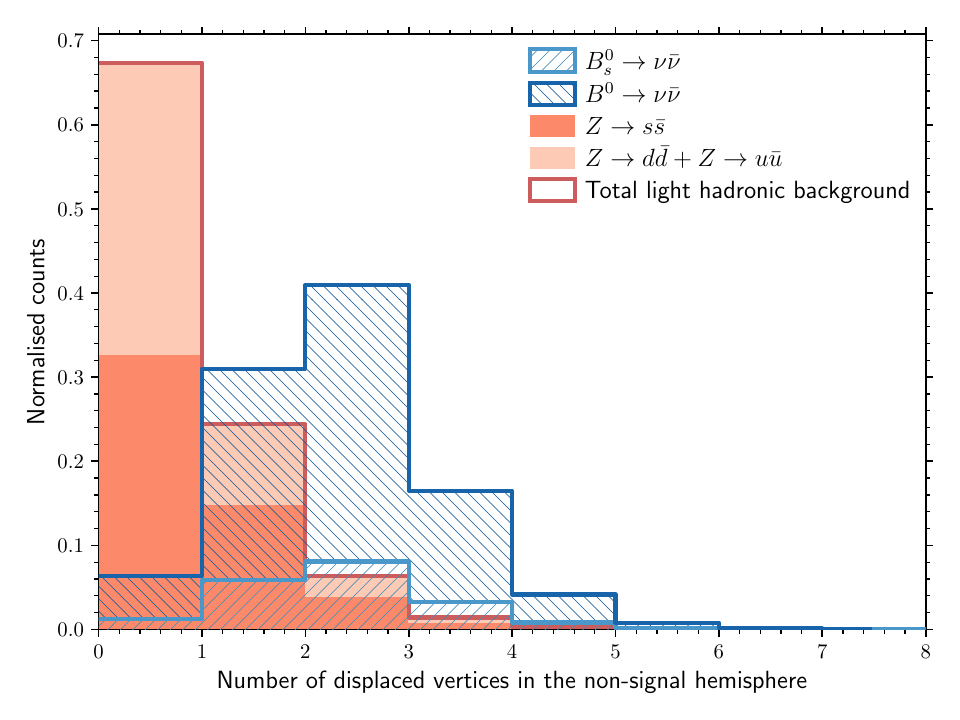} \\
    \includegraphics[width=0.49\textwidth]{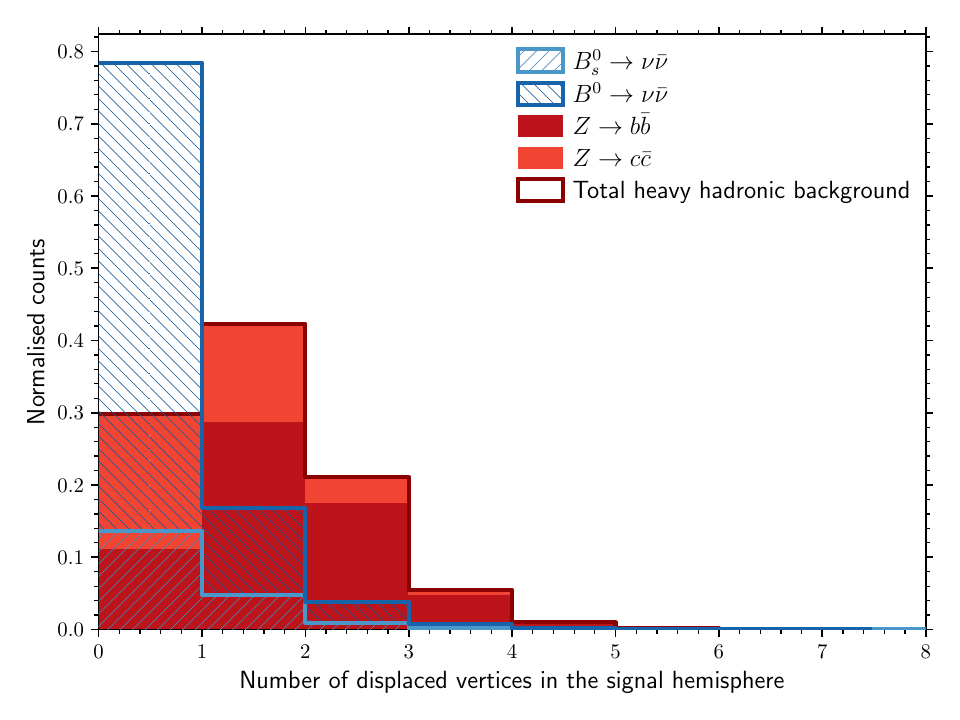}
    \includegraphics[width=0.49\textwidth]{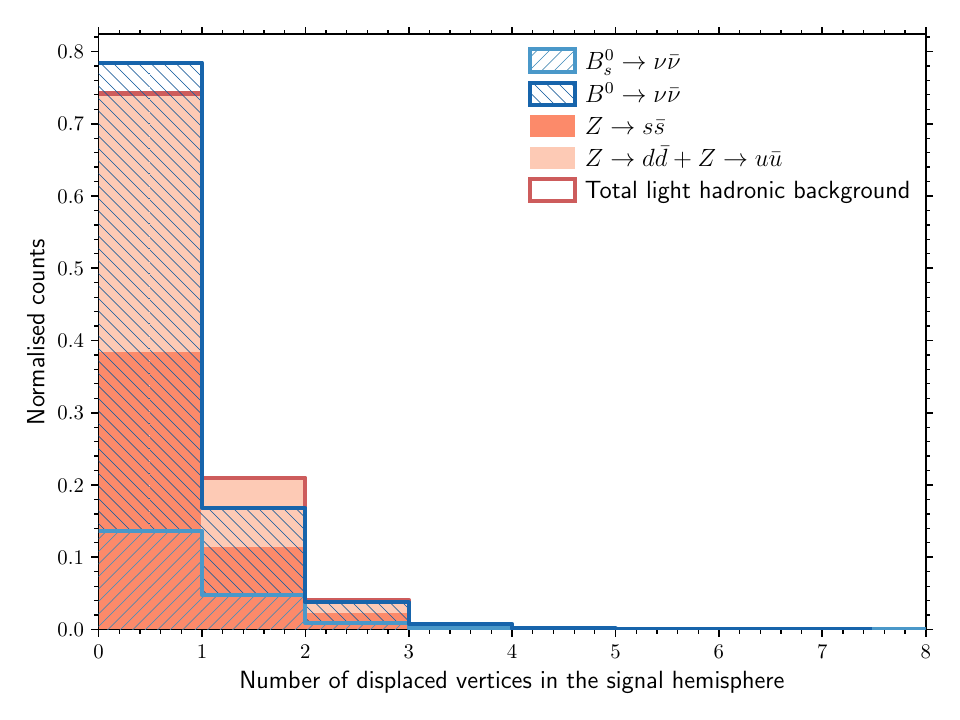}
    \caption{A demonstration of the different characteristics in the signal (bottom) and non-signal (top) hemispheres for signal (blue), heavy and light hadronic backgrounds (red). The number of displaced vertices in a given hemisphere for heavy (left) and light (right) hadronic backgrounds show signal (non-signal) hemisphere variables are powerful at rejecting heavy (light) hadronic backgrounds.}
    \label{fig:multiclass_inputs}
\end{figure}

The background from heavy hadrons, \Zbb and \Zcc, also has these characteristics, and thus looks similar to the signal on the opposite side, but does not have the characteristic large missing energy and low multiplicity on the signal side. 
These heavy backgrounds will also normally have displaced vertices and tracks on the signal side as well as the opposite side.
Conversely, the background from light hadrons, \Zss, \Zdd and \Zuu, looks more similar to the signal on the signal side, but not on the opposite side.
Subsequently, variables based on the signal hemisphere information tend to be better at rejecting the heavy background, whereas variables based on the non-signal hemisphere are better at rejecting the light background.
Figure~\ref{fig:multiclass_inputs} shows two representative input variables that demonstrate this behaviour, the number of displaced vertices reconstructed in each hemisphere.
There are several other variables which also exhibit this behaviour which motivates the use of a multiclass classifier that has three outputs, interpreted as the probability a given event is signal, heavy background or light background.
The raw output scores of the classifier have the \texttt{softmax} function applied such that the sum of the three probabilities for any given event is unity.

The classifier is trained using the \texttt{XGBoost} package~\cite{xgb} on the following set of 49 input variables:

\begin{itemize}
    \item The charged and neutral reconstructed energy in each hemisphere.
    \item The charged and neutral reconstructed particle multiplicity in each hemisphere.
    \item The total number of reconstructed tracks and vertices.
    \item The $x$, $y$ and $z$ components of the total visible momentum in the event.
    \item The fit quality and invariant mass of the PV.
    \item The $x$, $y$ and $z$ components of the thrust vector and the quality of thrust fit.
    \item The total number of displaced tracks in the signal and non-signal hemispheres.
    \item The number of tracks from the DV on the signal side with the largest number of tracks, and the non-signal side equivalent.
    \item The minimum and maximum distance between the PV and any DV in both the signal and non-signal sides.
    \item The maximum radial and longitudinal impact parameter (and impact parameter significance) for any track in both the signal and non-signal sides.
    \item The scalar momentum of the highest momentum charged track on the signal side and the non-signal side (and a flag for whether those candidates originated from the PV or not).
    \item The average of the cosine of the angle between each reconstructed track and the thrust vector in the signal and non-signal hemispheres.
    \item The minimum and maximum value of the angle between the thrust vector and the momentum of any vertex, in each hemisphere.
\end{itemize}

The BDT is trained using three classes of events: signal, heavy background and light background. 
The samples contain approximately 1M signal events (500K each of \bdinv and \bsinv), 1M heavy background events (500K each of \Zbb and \Zcc) and 1M light background events (500K each of \Zss and \Zud) which pass the preselection requirements described in Sec.~\ref{sec:preselection}.
These are split into a training sample, a test sample to check for overtraining and a validation sample which is used for hyperparameter optimisation, with a 75:12.5:12.5 split.
Events are appropriately weighted according to their known production fraction and preselection efficiency, so that the sum of weights for a given sample is proportional to the number of expected events for that sample.
Before the training, events are additionally given importance weights to ensure that the sum of weights for each of the three classes is the same. 

\begin{figure}
    \centering
    \includegraphics[width=0.49\textwidth]{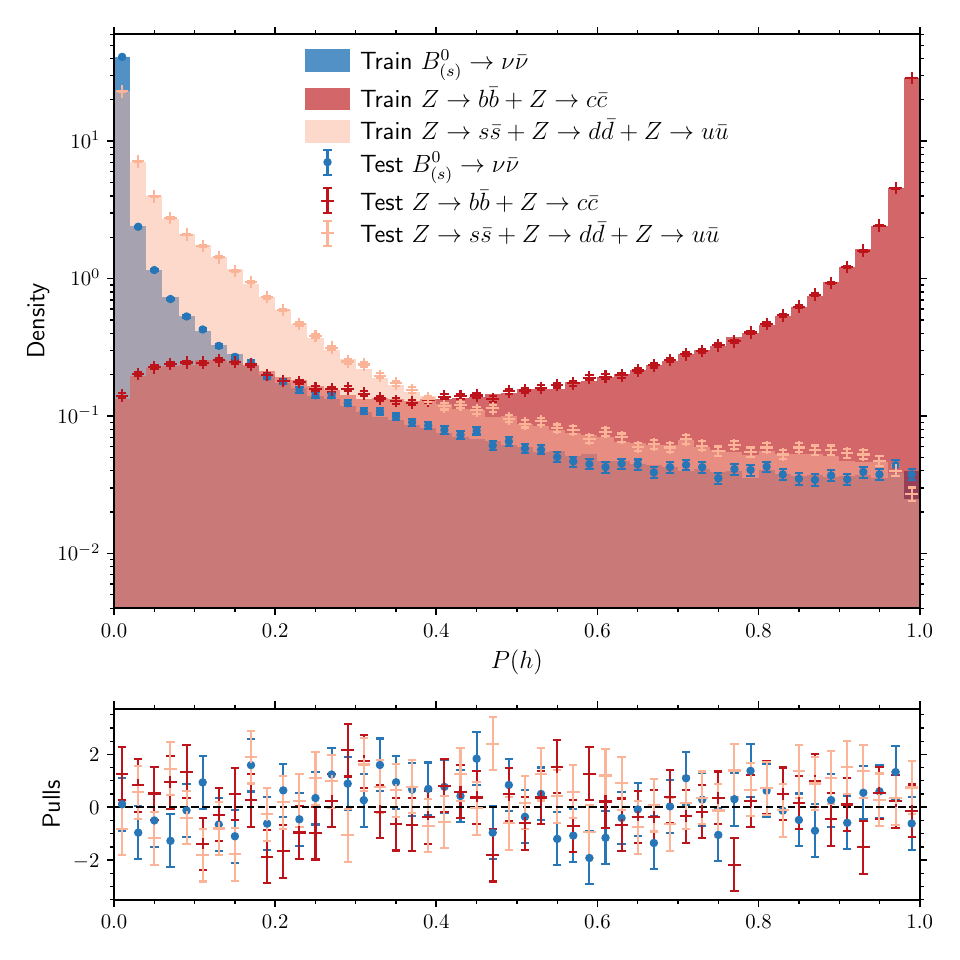}
    \includegraphics[width=0.49\textwidth]{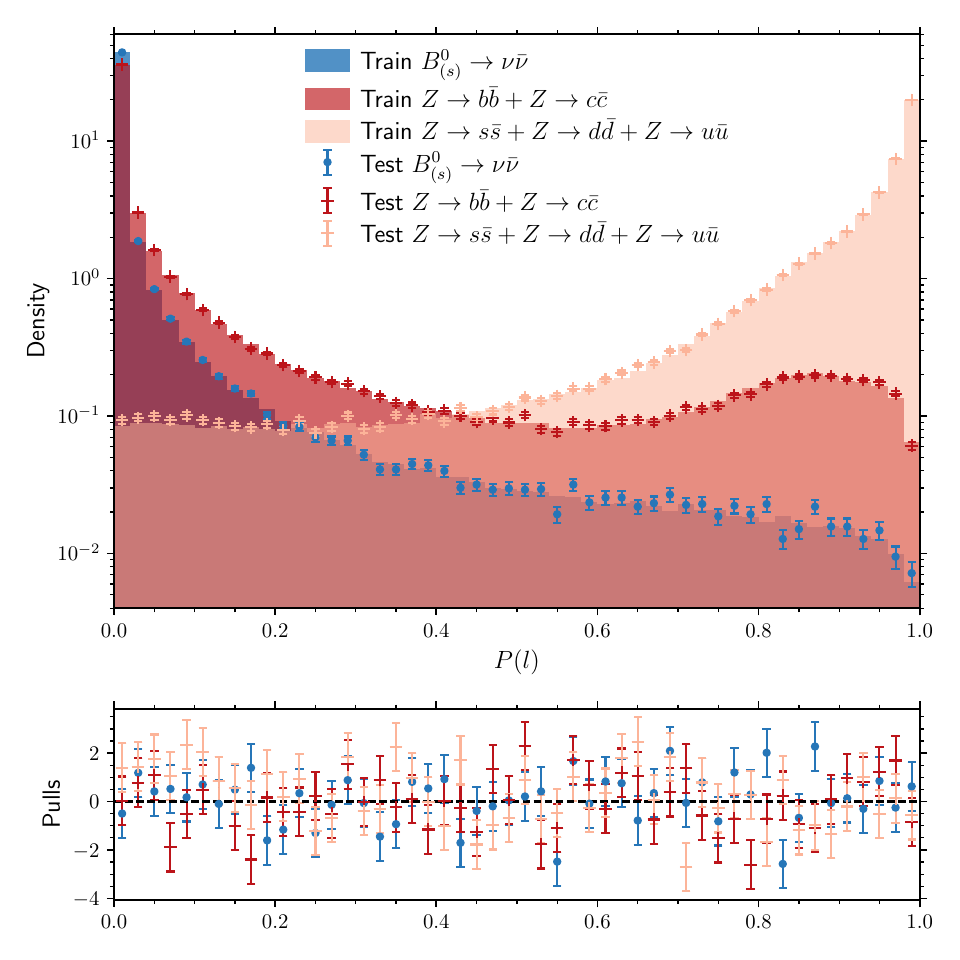}
    \includegraphics[width=0.49\textwidth]{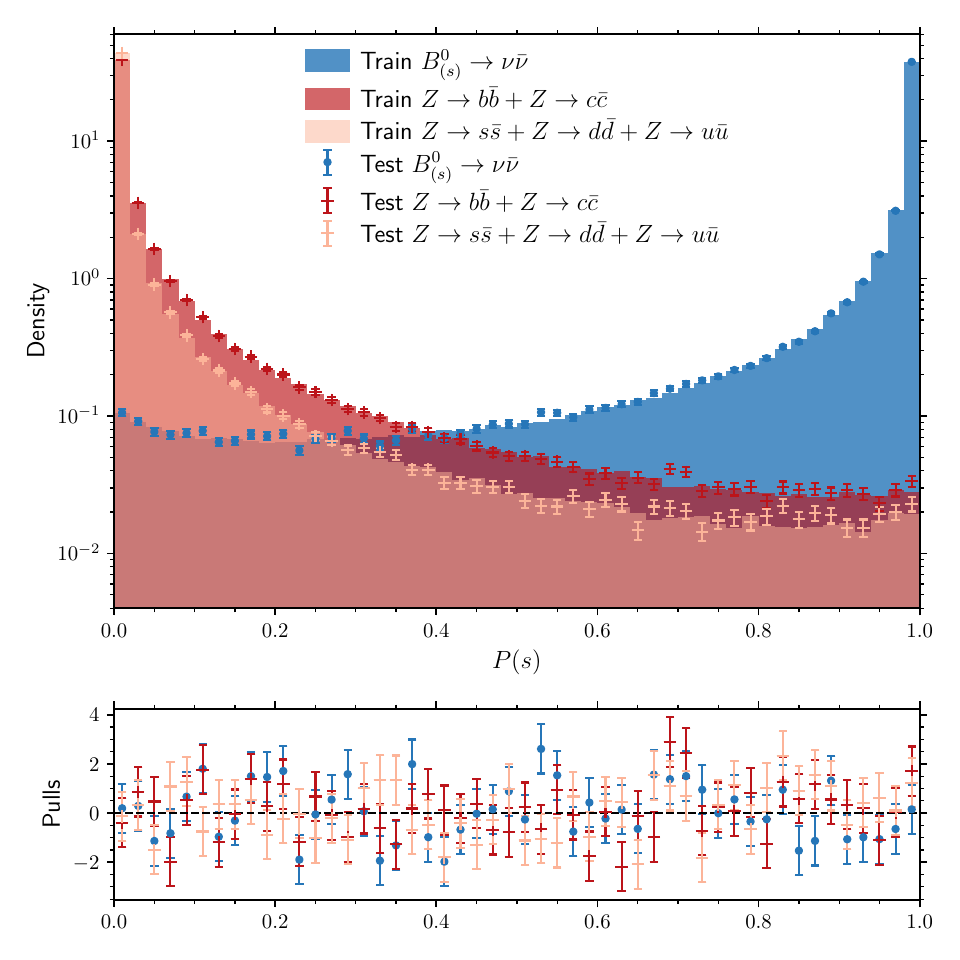}
    \caption{Distributions of the three BDT classifier outputs, interpreted as the probabilities an event is classified as heavy background (top left), light background (top right) or signal (bottom) for the three training samples: signal (blue), heavy background (dark red), light background (light red). The training sample is shown as the filled histogram and the test sample as the points with error bars. The observed agreement between test and train samples is consistent with no overtraining.}
    \label{fig:bdt_outputs}
\end{figure}

The classifier hyperparameters are optimised using the categorical-cross-entropy on the validation sample, balancing performance against overtraining.
The optimal hyperparameters chosen are 400 estimators, a learning rate of 0.1, a maximum depth of 4 and early stopping of 10 rounds.
The output of the three BDT scores, interpreted as the probability an event is classified as signal, heavy background or light background are shown in Fig.~\ref{fig:bdt_outputs}.
It should be noted that, due to implementation of the \texttt{softmax} function on the raw output scores of the classifier, for any given event the three output scores will sum to one, thus only two of the three output distributions are independent. 

\subsection{Background from leptonic $Z$ decays}
\label{sec:ztautau}

\begin{figure}
    \centering
    \includegraphics[width=0.49\textwidth]{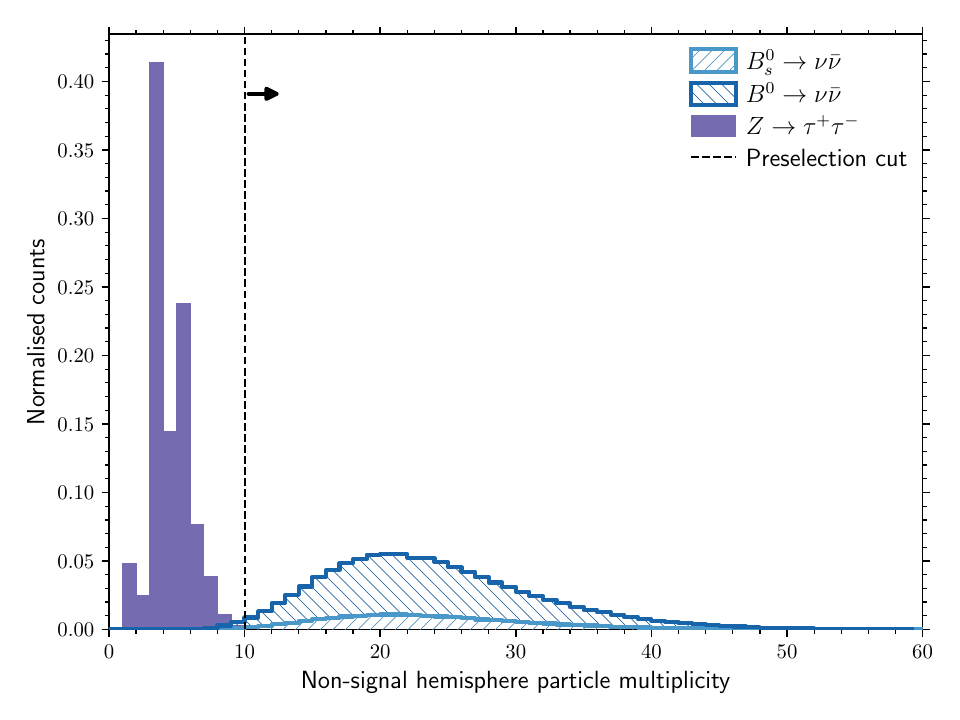}
    \caption{The total reconstructed particle multiplicity in the non-signal hemisphere for \Ztautau background (purple) and \binv (blue).}
    \label{fig:ztautau_bkg}
\end{figure}

Background contributions from \Zee and \Zmumu are reduced to entirely negligible levels by the preselection requirements described in Sec.~\ref{sec:preselection}.
They are reduced primarily by the requirement of no reconstructed muons or electrons on the signal side, which has almost zero probability for both \Zee and \Zmumu.
Contributions from \Ztautau are also effectively removed by the requirement that the non-signal side (charged and neutral) particle multiplicity is larger than ten.
The distribution for signal and \Ztautau background events in this variable is shown in Fig.~\ref{fig:ztautau_bkg}.
The preselection requirement efficiency on simulated \Ztautau events is $\sim (3\times 10^{-3})\%$ and the entire 100M sample available to us is removed after further requirements on the BDT outputs.
In reality, there may be some extreme cases in which \Ztautau events do pass these requirements, however there are several other variables that future work could exploit to separate \Ztautau background from the signal. 
These include, for example, compatibility of tracks to the beamspot, the quality of the thrust axis fit, the total number of vertices and tracks,  and impact parameter variables on the non-signal side.
These could be deployed in a multivariate classifier which specifically targets the \Ztautau background to further boost the performance of the \Ztautau rejection.

\subsection{Background from \BuorBc\to~\ellp\neul decays}\label{sec:b2lnu}

Missing the charged lepton when reconstructing \BuorBc\to~\ellp\neul decays would produce background events which look signal-like to our selection. 
The dominant of these helicity suppressed backgrounds are \mbox{$\BuorBc\to\taup\neut$} decays, which 
account for $\sim20\%$ of the inclusive \Zbb background events which pass the full selection.

To further study these modes, a cocktail of exclusive MC samples, including contributions from $\BuorBc \to \taup \neut$ (where the $\taup$ decays hadronically or leptonically) and $\BuorBc \to \mup \neum$ were produced.
Decays of \mbox{$\Bu\to~\taup\neut$} where the $\taup$ undergoes a one-prong decay were found to dominate, with the signal side charged lepton veto suppressing one-prong $\ep$ and $\mup$ final states by a factor of $\sim3.5$ compared to the one-prong $\pip$.   
In a future analysis, this one-prong pion background could be further reduced using a veto on displaced isolated signal-side charged hadrons. Furthermore, the high granularity calorimeter of the IDEA detector~\cite{IDEAStudyGroup:2025gbt} could be utilised in sophisticated $\tau$-reconstruction algorithms~\cite{Giagu:2022gmq} to provide stronger rejection of  \mbox{$\Bu\to~\taup\neut$} background where \mbox{\taup\to~\hp\neutb} with one or more neutral \piz.

\subsection{Other backgrounds}

In future analyses, more targeted selection requirements, such as those exploiting PID, isolation criteria, and track separation from the PV, could be employed to further suppress contributions still present in the inclusive background samples after the full  selection is applied.
These backgrounds are primarily due to \B-meson decays to neutral final states, for example \mbox{\BdorBs\to~\Kz\piz}, which have branching fractions of $\mathcal{O}(10^{-5})-\mathcal{O}(10^{-6})$; semileptonic \B decays in which the hadronic component subsequently decays (semi)leptonically or to neutral particles; and radiative \mbox{\bquark\to~\squark\g} transitions such as \mbox{\Lb\to~\Lz\g}, with a branching fraction of $\sim7\times10^{-6}$. 
Similar strategies would also further suppress residual charm backgrounds, including \mbox{\Dz\to~\Kz\piz} and \mbox{\Dp\to~\Kz\pip} decays with branching fractions of roughly $2-3\%$; the semileptonic modes \mbox{\Dp\to~\ep\neue\Kzb} and \mbox{\Dp\to~\mup\neum\Kzb}, each with branching fractions of about $9\%$; and \mbox{\Dsp\to~\taup\neut} decays, which have a branching fraction of approximately $5\%$.
In the latter case, one-prong tau decays dominate, mirroring the behaviour in the \mbox{\BuorBc\to~\taup\neut} background. 
Furthermore, approximately $90\%$ of the light hadronic background events, from the \Zss and \Zud inclusive MC samples that pass the full selection, contain a \cquark\cquarkbar or \bquark\bquarkbar quark pair, which hadronises to form two heavy \cquark  or \bquark hadrons on the non-signal side. 
Dedicated selection  requirements designed to target this characteristic topology  could therefore be used to further reduce their contribution.

\subsection{Sensitivity estimate} \label{sec:sensitivity}

To obtain an estimate for the overall sensitivity of a search for \binvisible decays at FCC-ee, optimal BDT cuts were first selected for each value of interest of the combined signal branching fraction $\BF(\binvisible)$. Requirements were placed on two of the BDT output scores, namely the probabilities of a given event being heavy background, $P(h)$, and light background, $P(l)$. This enables the multiclass BDT's power to be harnessed to find a more effective cut point compared to a simple cut on the probability of an event being signal, $P(s) = 1 - P(h) - P(l)$.

The figure of merit used for this optimisation was the simple sensitivity of a single-bin counting experiment, defined as 
\begin{equation}
\label{eq:naiveFOM}
    \text{FoM} = \frac{S}{\sqrt{S+B}}.
\end{equation}
The signal expectation, $S$, is computed using 
\begin{equation}
    \label{eq:NS}
    S = \displaystyle\sum_{k \in \{\Bd, \Bs\}}2 \, N_Z \,  \BF(\Z\to\bbbar) \, f_{k} \, \BF(\binvisible) \, \epsilon^{s}_{k},  
\end{equation}
where $N_Z$ is the number of $Z$ bosons produced, the factor of two accounts for the two \bquark quarks produced per \Zbb decay, $f_k$ is the production fraction for the \bquark quark to hadronise into the relevant \bquark hadron (\Bd, \Bs), $\epsilon^{s}_k$ is the relevant efficiency of the full selection, and $\BF(\binvisible)$ is the hypothesised signal branching fraction.\footnote{We assume that the branching fractions for \bdinvisible and \bsinvisible are equal. This is not exactly the case for the SM given the different CKM elements involved between a \Bd and \Bs penguin decay, and the small difference in kinematics due to the \Bd-\Bs mass difference.}
The background expectation, $B$, is computed using  
\begin{equation}
    \label{eq:NB}
    B = \displaystyle\sum_{f \in \{\bbbar, \ccbar, \ssbar, \ud\}} N_Z \, \BF(\Z \to f) \, \epsilon^{b}_{f} ,
\end{equation}
where $\BF(\Z\to f)$ are the relevant branching fractions for $Z\to\text{hadrons}$ (either \bbbar, \ccbar, \ssbar or \ud) and $\epsilon^{b}_{f}$ is the full selection efficiency of the relevant background component.

Throughout this study we assume the following values of the parameters in Eqs.~\eqref{eq:NS} and~\eqref{eq:NB}:
\begin{itemize}
    \item $N_Z=6\times 10^{12}$, the number of \Z bosons produced across all experiments during the entire Tera--\Z run at FCC-ee.
    \item The production fraction of $\B$ mesons from \Zbb decays are $f_\Bd=0.408(7)$ and \mbox{$f_\Bs=0.100(8)$} \cite{Amhis_2021,ParticleDataGroup:2024cfk}.
    \item The $\Z\to\text{hadrons}$ branching fractions are $\BF(\Zbb) = 0.1512(5)$, $\BF(\Zcc)=0.1203(21)$, $\BF(\Zss)=0.1584(60)$ and $\BF(\Zud)=0.2701(136)$~\cite{ParticleDataGroup:2024cfk}.
\end{itemize}

\begin{figure}
    \centering
    \begin{minipage}[t]{0.29\textwidth}
        \centering
        \begin{subfigure}[t]{\textwidth}
            \centering
            \caption{\bdinv}
            \includegraphics[width=\textwidth]{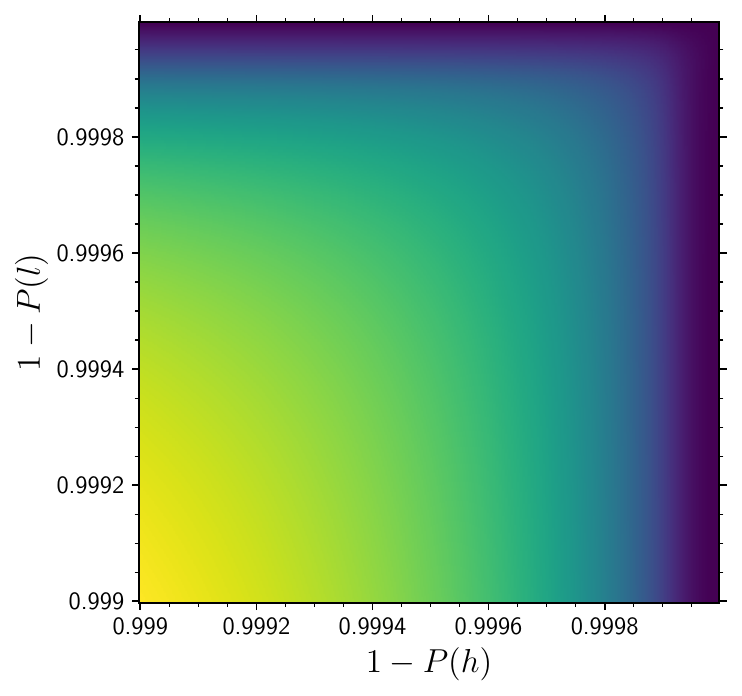}
        \end{subfigure}
        \vspace{1em}
        \begin{subfigure}[t]{\textwidth}
            \centering
            \caption{\bsinv}
            \includegraphics[width=\textwidth]{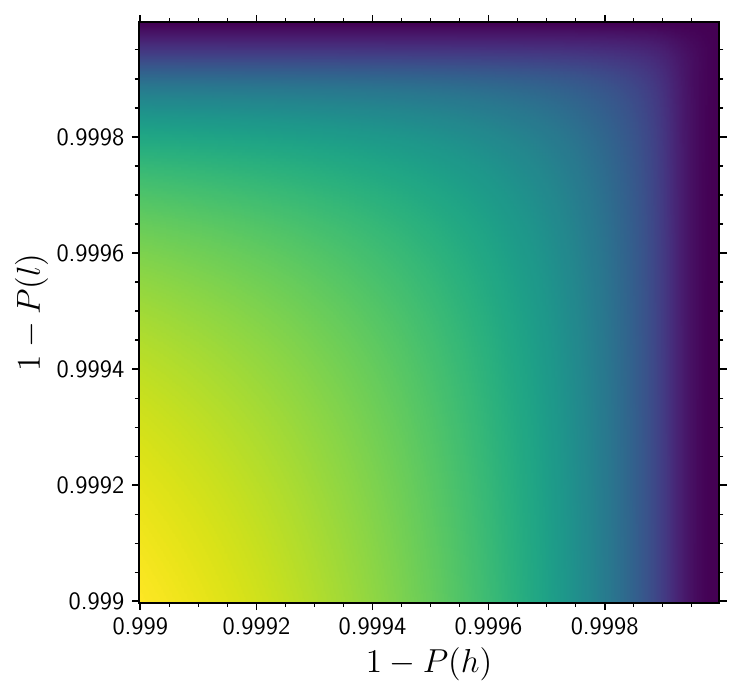}
        \end{subfigure}
    \end{minipage}%
    \hfill
    \begin{minipage}[t]{0.29\textwidth}
        \centering
        \begin{subfigure}[t]{\textwidth}
            \centering
            \caption{\Zbb}
            \includegraphics[width=\textwidth]{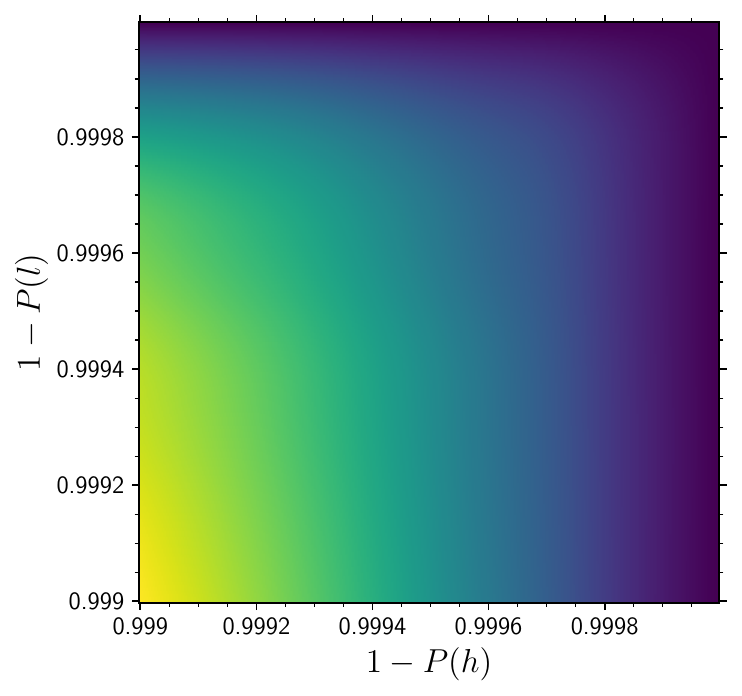}
        \end{subfigure}
        \vspace{1em}
        \begin{subfigure}[t]{\textwidth}
            \centering
            \caption{\Zss}
            \includegraphics[width=\textwidth]{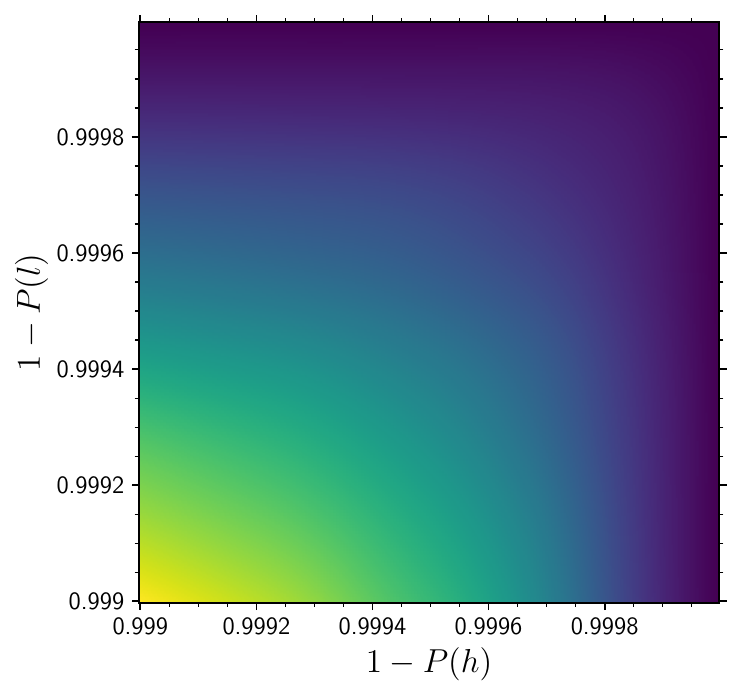}
        \end{subfigure}
    \end{minipage}%
    \hfill
    \begin{minipage}[t]{0.29\textwidth}
        \centering
        \begin{subfigure}[t]{\textwidth}
            \centering
            \caption{\Zcc}
            \includegraphics[width=\textwidth]{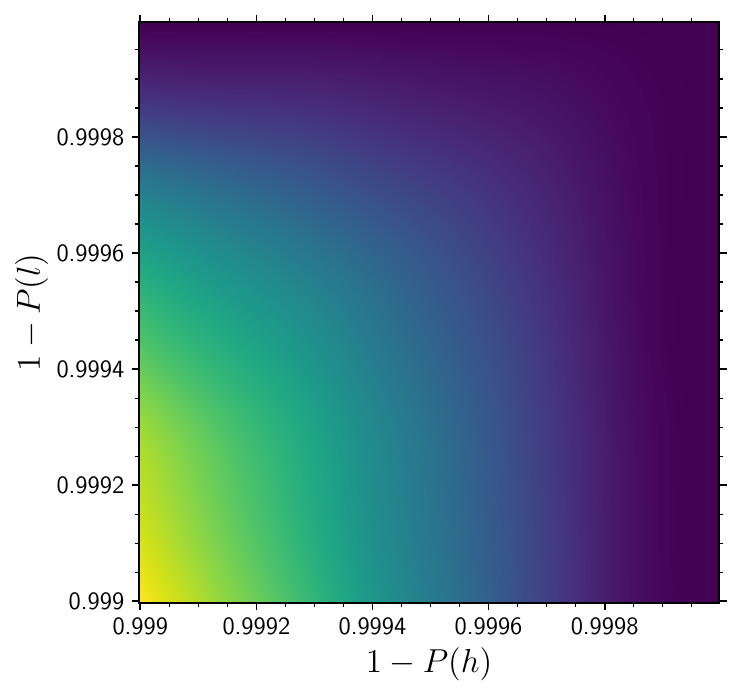}
        \end{subfigure}
        \vspace{1em}
        \begin{subfigure}[t]{\textwidth}
            \centering
            \caption{\Zud}
            \includegraphics[width=\textwidth]{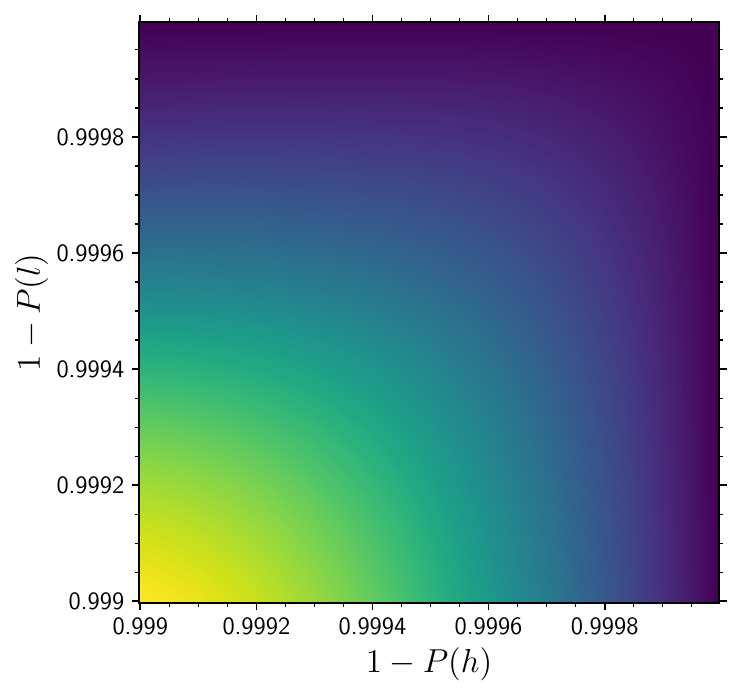}
        \end{subfigure}
    \end{minipage}%
    \hfill
    \raisebox{-0.2575\textheight}{%
        \begin{minipage}[c]{0.1\textwidth}
            \centering
            \includegraphics[width=\textwidth]{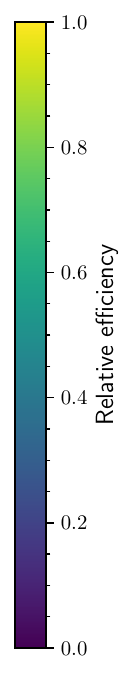}
        \end{minipage}
    }

    \caption{Interpolated maps of the fraction of MC events remaining across a grid of cuts on the BDT output scores parameterising the probability of an event being not-heavy background, $1-P(h)$, and not-light background, $1-P(l)$.}
    \label{fig:N_interp}
\end{figure}

To obtain selection efficiencies for the signal and background samples, maps of the number of events remaining in each MC sample were quadratically interpolated over a coarse two-dimensional grid of cuts on $1-P(h)$ and $1-P(l)$, as shown in Fig.~\ref{fig:N_interp}. 
Interpolation was performed due to the low MC statistics remaining after the tight BDT cuts necessary to sufficiently reduce the background. 
From the interpolated maps, the total efficiency of the full selection, and therefore the figure of merit in Eq.~\eqref{eq:naiveFOM}, was calculated for any pair of cuts on  $1-P(h)$ and $1-P(l)$, and a grid search was performed to find the optimum cut for a given signal branching fraction.  
Assuming an expected signal branching fraction of $7.6 \times 10^{-9}$, optimum cuts give the selection efficiencies detailed in Table~\ref{tab:full_efficiencies}.

\setlength{\tabcolsep}{2pt}
\renewcommand{\arraystretch}{1.2}
\begin{table}
    \centering
    \begin{tabular}{l r c l}
        \textbf{Sample} & \multicolumn{3}{c}{\textbf{Efficiency (\%)}} \\
        \hline
        \bdinv       & $19.57$ & $\pm$ & $0.03$ \\
        \bsinv       & $18.07$ & $\pm$ & $0.03$ \\
        \hline
        \Zbb         & ($5.8$ & $\pm$ & $0.4$)$\times 10^{-5}$ \\
        \Zcc         & ($2.5$ & $\pm$ & $0.2$)$\times 10^{-5}$ \\
        \Zss         & ($3.5$ & $\pm$ & $0.3$)$\times 10^{-5}$ \\
        \Zud$\qquad$ & ($3.8$ & $\pm$ & $0.9$)$\times 10^{-6}$ \\

    \end{tabular}
    \caption{Total selection efficiencies for each of the samples used in this analysis. The sample labelled as \Zbb includes only background candidates.  For each of the three leptonic samples  (\Ztautau, \Zmumu and \Zee), all 100 million MC events were completely removed by the full selection.}
    \label{tab:full_efficiencies}
\end{table}

Three different sensitivity estimates are made, relying on various levels of assumption. The first is a straightforward estimate based on a single-bin counting experiment, given in Eq.~\eqref{eq:naiveFOM}, which assumes no systematic uncertainty on the signal and background expectations, $S$ and $B$, respectively. However, due to the very small anticipated signal branching fraction, tight cuts need to be placed on both BDT scores leading to potentially non-negligible systematic contributions to the background expectation. This prompted the use of a second estimate, defined as
\begin{equation}
    \label{eq:FOMinclerr}
    \frac{S}{\sqrt{S+B+\sigma_S^2+\sigma_B^2}},
\end{equation}
where $\sigma_S$ and $\sigma_B$ account for additional uncertainties in the estimates of $S$ and $B$, respectively. These include the uncertainty on the efficiency estimates, $\epsilon_{f/k}^{s/b}$, due to the finite size of the simulation samples computed using a symmetrised Wilson interval~\cite{wilson_efferr}, the uncertainty on the hadronic branching fractions, $\BF(\Z \to f)$ for $f\in \{\bbbar, \ccbar, \ssbar, \ud\}$, and the uncertainty on the fragmentation fractions $f_\BdorBs$.

\begin{figure}
    \centering
    \includegraphics[width=0.49\linewidth]{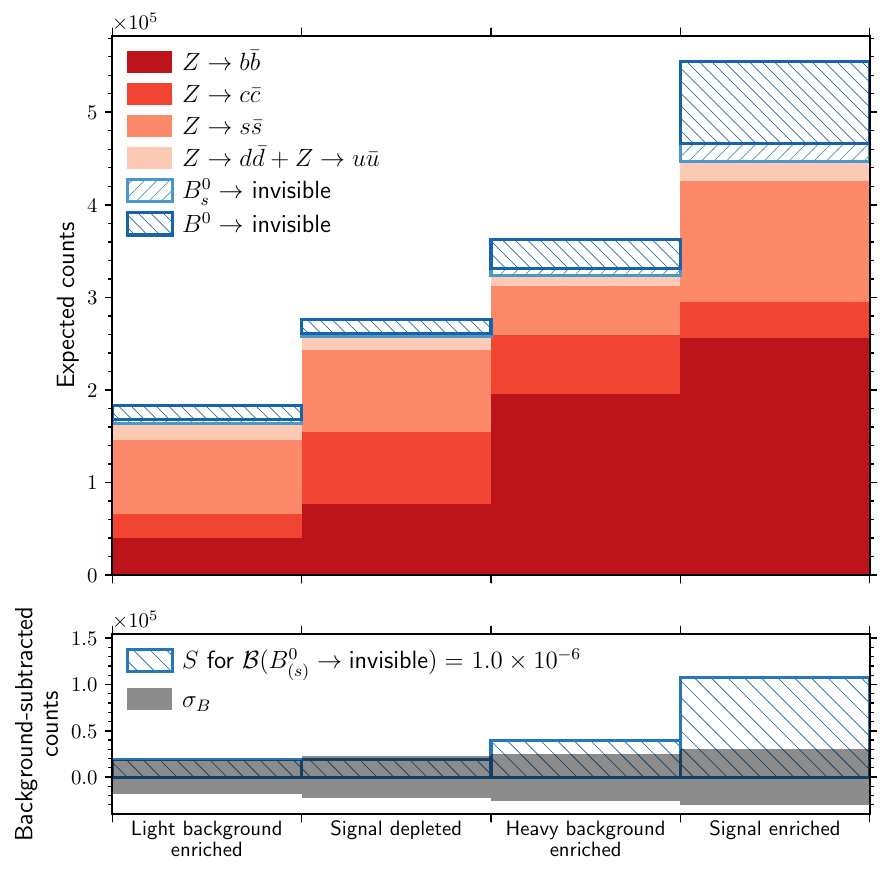}
    \includegraphics[width=0.49\linewidth]{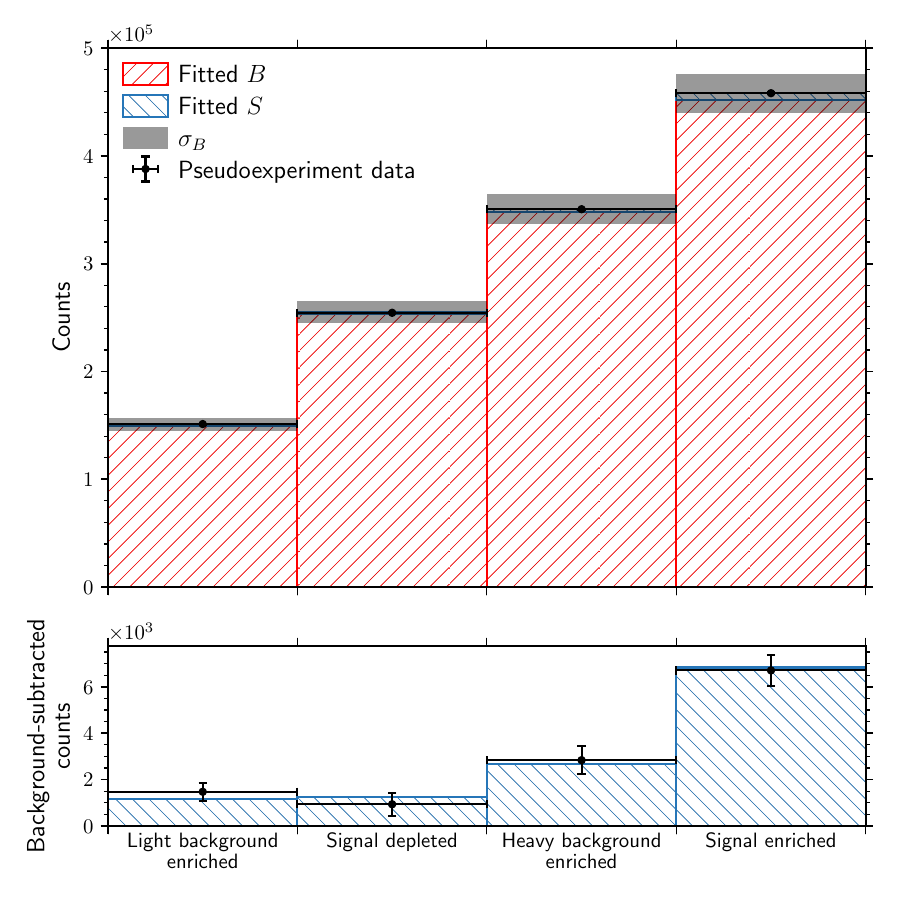}
    \caption{Left: the distribution of signal and background proxies from simulation samples, binned in the two-dimensional space of the BDT output variables $P(h)$ and $P(l)$, used to generate pseudoexperiments, for a signal branching fraction of $1 \times 10^{-6}$.
    Right: Example binned fit to pseudoexperiment data generated from Poisson distributions of signal and background expectations for a signal branching fraction of $7.1\times10^{-8}$.}
    \label{fig:finalbinning}
\end{figure}

This estimate still assumes a single-bin counting experiment. Some of the sensitivity lost by the inclusion of systematic contributions can be recovered by fitting to separate bins in the BDT outputs, motivating a third sensitivity estimate based on performing binned fits to ensembles of pseudoexperiments. 
After selecting an optimum set of BDT cuts,  the remaining parameter space in $1-P(h)$ and $1-P(l)$ was split into four equal bins.
An example is shown for a hypothesised signal branching fraction of $1\times 10^{-6}$ in Fig.~\ref{fig:finalbinning} (left).
Fits were performed to an ensemble of 10,000 pseudoexperiments, randomly sampled using Poisson statistics from the signal and background expectations in each bin. It was assumed that the shape of the binned distributions was known perfectly, with the overall size of the signal and background contributions allowed to float.
An example of one of the fit results is shown in Fig.~\ref{fig:finalbinning} (right).
A Gaussian constraint was used to account for the non-trivial systematic uncertainty, $\sigma_B$, which is assumed to be fully correlated across the bins.
The sensitivity was then computed using Wilks' theorem~\cite{Wilks:1938}, assuming that $\sqrt{-2 \Delta\ln{\mathcal{L}}}$ gives a suitable proxy for the significance, where $\Delta\ln{\mathcal{L}}$ is the difference in log-likelihood between the background-only fit (forcing $S=0$ across all bins) and the best fit (allowing $S$ to float).

\begin{figure}
        \centering \includegraphics[width=0.49\textwidth]{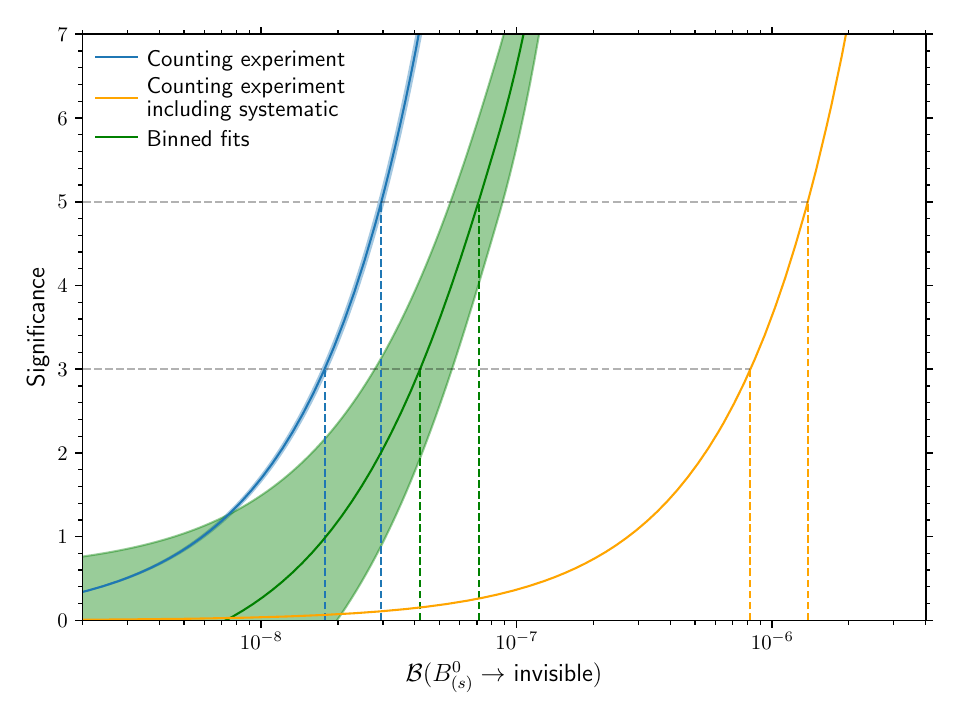}
    \includegraphics[width=0.49\textwidth]{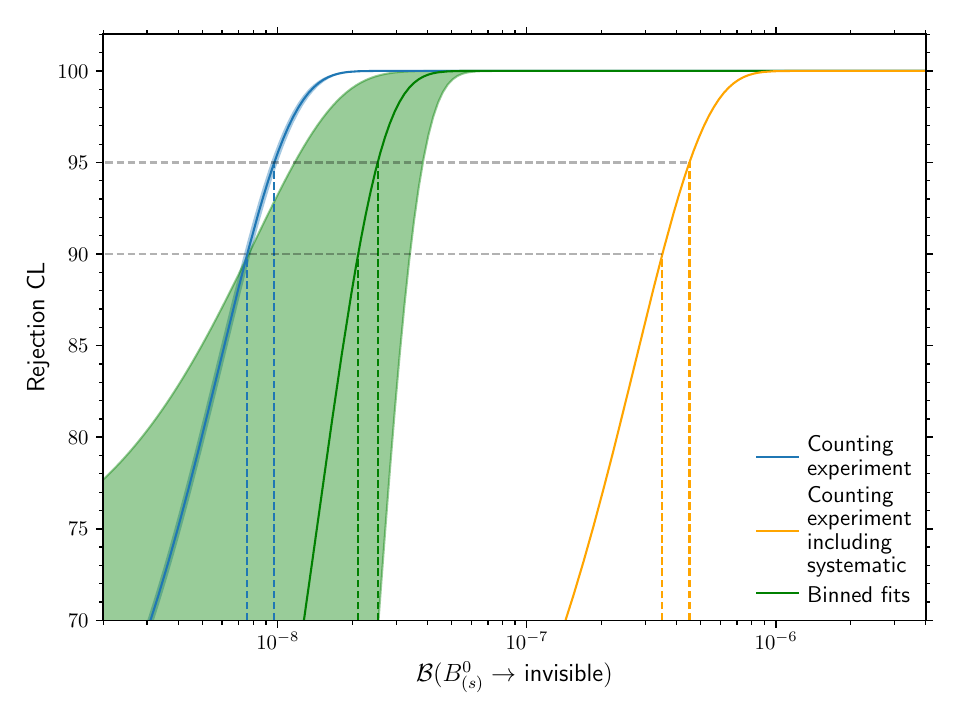}
    \caption{Anticipated constraints FCC-ee would be able to set on \binvisible decays at different branching fractions, depending on the assumptions made in the signal and background modelling. The green band is determined from the variation across the ensemble of pseudoexperiments and contains 68.3\% of the distribution.}
    \label{fig:significance}
\end{figure}

The three significance estimates are shown as a function of the hypothesised \binvisible branching fraction in Fig.~\ref{fig:significance}.
Table~\ref{tab:sensitivities} shows
estimates of the signal branching fraction which could be within discovery reach of, or excluded at, FCC-ee if it produces $6\times 10^{12}$ $Z$ bosons.
The values we extract using the binned fit methodology (green line in Fig.~\ref{fig:significance}) are consistent with the reinterpretation of ALEPH data presented in Ref.~\cite{Alonso-Alvarez:2023mgc}, once approximate scaling for the larger number of $Z$ bosons and improved efficiency of this analysis, have been included.

\setlength{\tabcolsep}{6pt}
\renewcommand{\arraystretch}{1.5}
\begin{table}
    \centering
    \resizebox{\textwidth}{!}{
    \begin{tabular}{l cccc}
         \textbf{Method}             & 90\% CL & 95\% CL & 3$\sigma$ &5$\sigma$ \\
        \hline
         Single-bin counting experiment  & $7.6\times10^{-9}$ & $9.7\times10^{-9}$ & $1.8\times10^{-8}$&$3.0\times10^{-8}$ \\
         Counting experiment including systematic & $3.5\times10^{-7}$ & $4.5\times10^{-7}$ & $8.2\times10^{-7}$&$1.4\times10^{-6}$ \\
         Binned fits to pseudoexperiments & $2.1\times10^{-8}$  & $2.5\times10^{-8}$ & $4.2\times10^{-8}$&$7.1\times10^{-8}$\\

    \end{tabular}
    }
    \caption{Expected \binvisible branching fractions probeable with $6\times 10^{12}$ $Z$ bosons, based on three different levels of assumption in the corresponding sensitivity estimate.}
    \label{tab:sensitivities}
\end{table}

\subsection{Separating the \Bd and \Bs signal}
\label{sec:separating_bd_and_bs}

The sensitivity estimates outlined in Sec.~\ref{sec:sensitivity} are given for some general \binvisible signal with a common branching fraction for \bdinvisible and \bsinvisible decays. This results in a combined sensitivity estimate, with the number of signal events of each type weighted by the ratio of \Bd to \Bs production fractions. 
A preliminary study was performed to assess the extent to which the two signals can be separated to allow separate limits to be set for invisible \Bd and \Bs decays. 
This requires accurate identification of the species of the signal-side \B meson, which decays invisibly, via its hadronisation partner(s). 
Due to the strangeness of the \Bs meson, separation is expected to be achievable by searching for reconstructible signal-side kaons (\Kpm, \KS). 
This study makes use of the underlying truth information from simulation and assumes perfect particle identification and vertex assignment.
Candidate \KS decays are selected by reconstructing charged (\pip\pim) and neutral ($\piz\piz \to \gamma\gamma\gamma\gamma$) pion pairs whose true parent or grandparent is a \KS.

\begin{figure}
    \centering \includegraphics[width=0.49\textwidth]{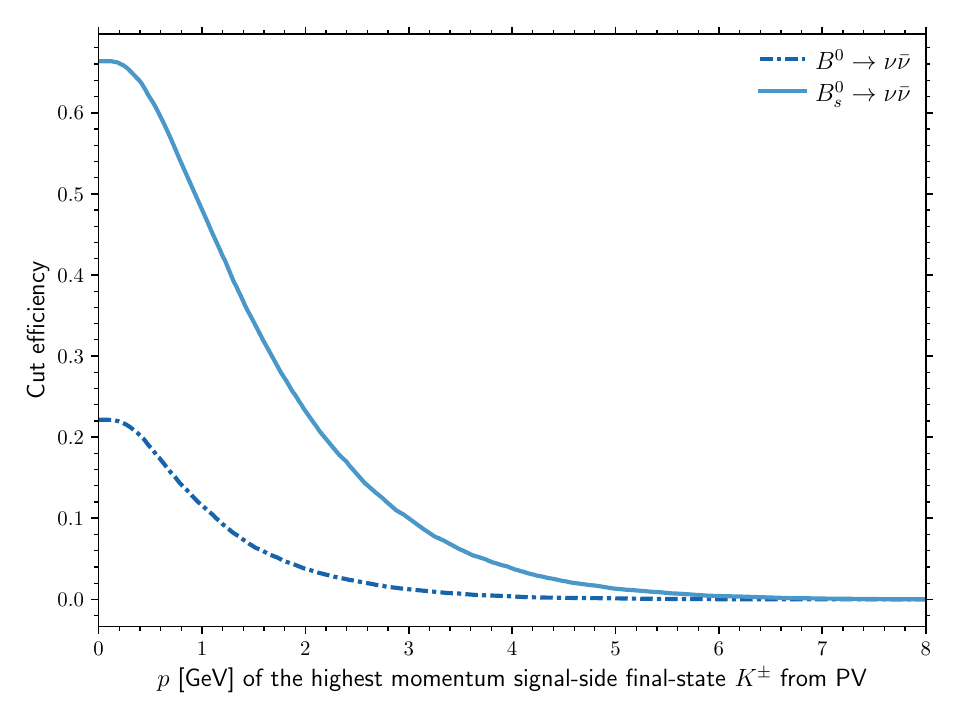}
    \includegraphics[width=0.49\textwidth]{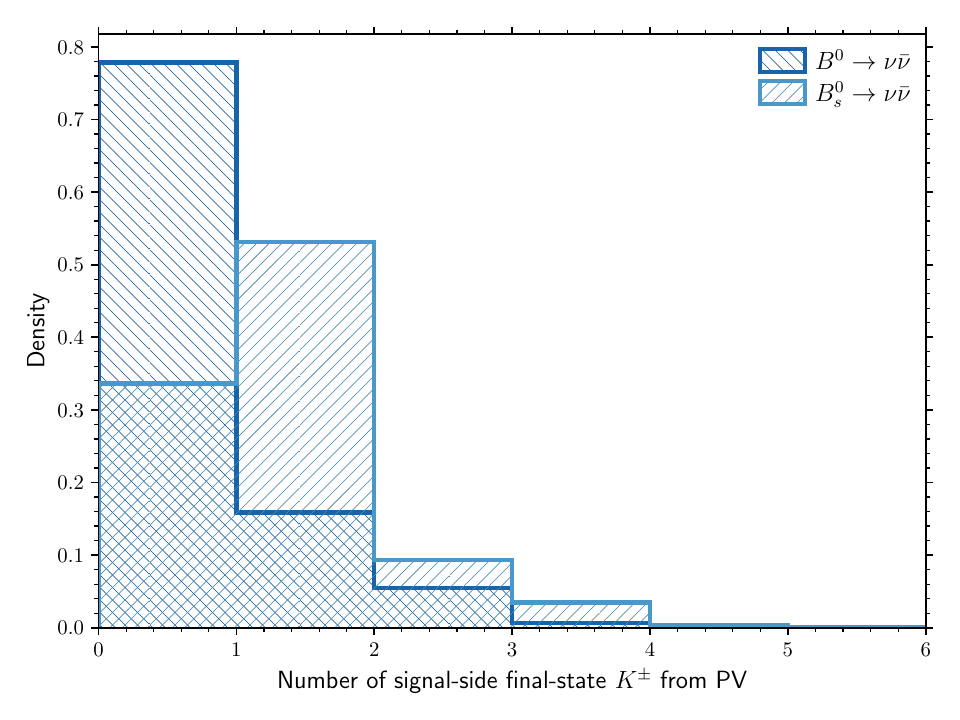}
    
    \includegraphics[width=0.49\textwidth]{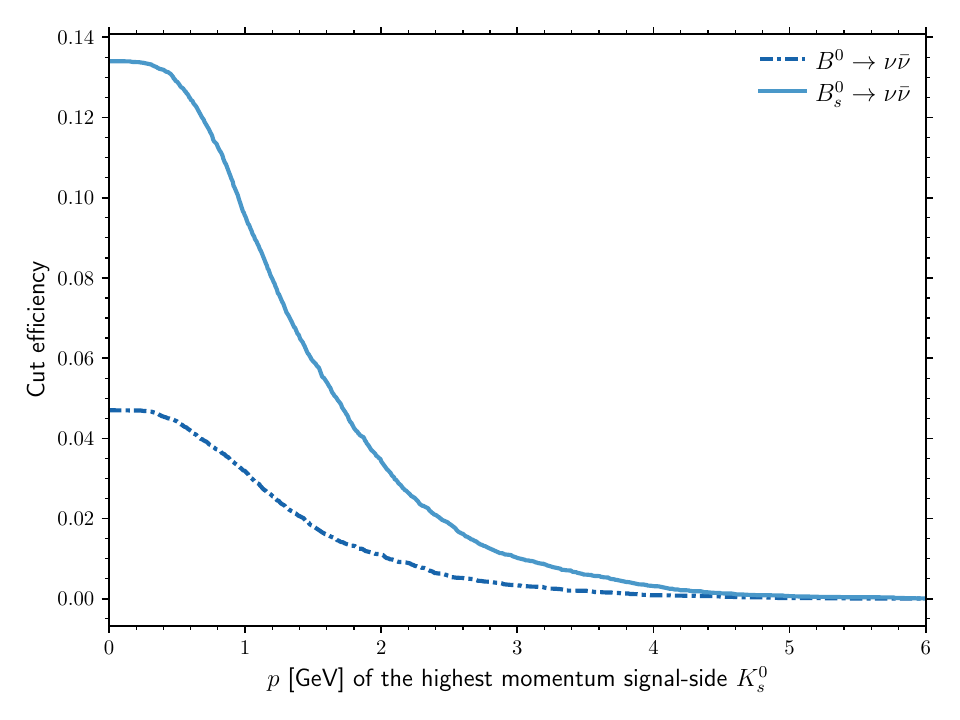}
    \includegraphics[width=0.49\textwidth]{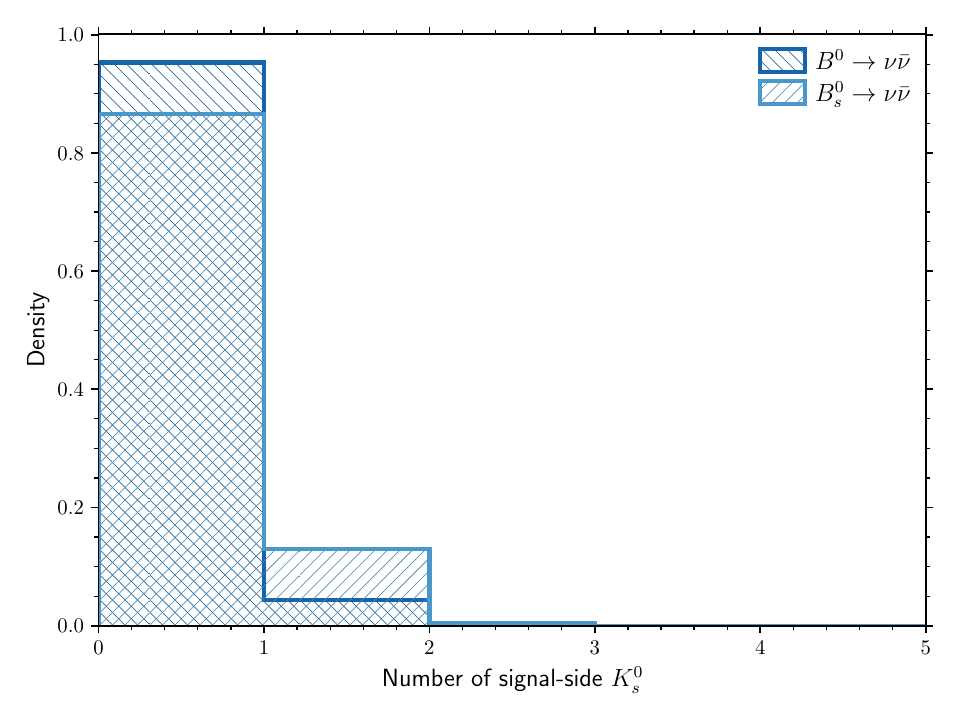}
    \caption{Estimated \Bd-\Bs separation achievable by selecting on the momentum of the highest momenta (left) and number (right) of prompt final-state \Kpm (top) and fully reconstructed intermediate \KS (bottom).}
    \label{fig:BsBd_separation}
\end{figure}

Figure \ref{fig:BsBd_separation} (right) shows the distribution of the numbers of charged and neutral kaon candidates in the signal hemisphere for the \Bd and \Bs signal. 
Figure \ref{fig:BsBd_separation} (left) shows the tagging efficiency as a function of a requirement on the scalar momentum of the hardest kaon candidate in the signal hemisphere, using transverse momentum gives similar results.
A retention rate of around 66\% of \Bs events whilst rejecting around 78\% of \Bd events is possible when requiring at least one reconstructed prompt charged kaon in the signal hemisphere.

These studies suggest that some separation of \Bd and \Bs samples could be achieved, however efficiencies of such a selection would be expected to impact the limits set. Limited further sensitivity could be gained by cutting on the prompt \Kpm or \KS momentum seen in Fig.~\ref{fig:BsBd_separation}.
Exploitation of sophisticated tagging methods with modern machine-learning architectures would likely help to further improve performance~\cite{Blekman:2024wyf}.

\subsection{Potential systematic effects}
\label{sec:systematics}

A real-life study will also incur a variety of systematic effects that will need to be considered. 
In terms of the analysis itself, the MC tuning and sample size will be important considerations, as will detailed studies of the dominant background contributions and detector effects.
In terms of the calculation of the branching fraction from the signal yield, Eq.~\eqref{eq:NS}, there will be additional sources of systematic uncertainty related to knowledge of the selection efficiencies, hadronisation fragmentation and production fractions, decay multiplicities, and related branching fractions that will need to be considered. 
Many of these quantities remain best measured by the LEP experiments, although with FCC-ee the precision of these measurements will substantially improve, thus reducing the systematic impact on this analysis.

The most significant external measurement systematics on this analysis arise from knowledge of the $Z\to\bbbar$ branching fraction and the \bquark-quark fragmentation fractions, $f_B$. 
The former is already known from LEP to 3 per mille precision~\cite{PDG2022} and with likely improvements from FCC-ee measurements will not have a significant impact on the precision of this analysis.
The latter, however, is currently only known to $\sim 2\%$ precision for the \Bd ($\sim 8\%$ for the \Bs)~\cite{PDG2022} meaning a potentially significant systematic impact on this analysis if further enhancements are not performed at FCC-ee. 
It is expected that FCC-ee itself will be able to improve knowledge of the fragmentation fractions by an order of magnitude or more, reducing the systematic impact to the same order as the statistical precision.

The dominant uncertainty in our analysis arises from the limited size of the MC simulation samples. 
The sensitivity estimates in Fig.~\ref{fig:significance} and Table~\ref{tab:sensitivities} have very limited dependence on any MC truth information.
No PID information is used which would likely help to improve the performance although would induce further systematic uncertainties. The multiclass BDT includes information relating to secondary vertex positions and fit qualities. 
In our analysis framework the secondary vertex fits are seeded by associated tracks to their true origin vertex. 
This assumption in principle enhances the sensitivity with respect to the real-life situation, however other studies have shown that if the vertex resolution of IDEA performs as expected, $\mathcal{O}(10 \mum)$, then the impact is negligible~\cite{Amhis:2023mpj}.

\begin{figure}
        \centering \includegraphics[width=0.49\textwidth]{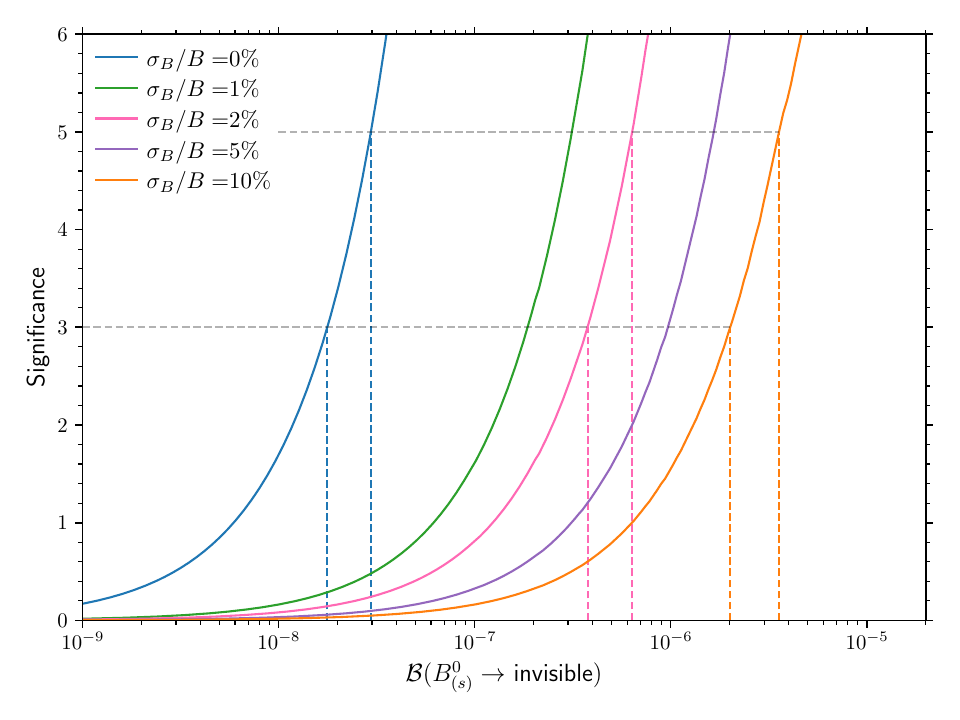}
    \includegraphics[width=0.49\textwidth]{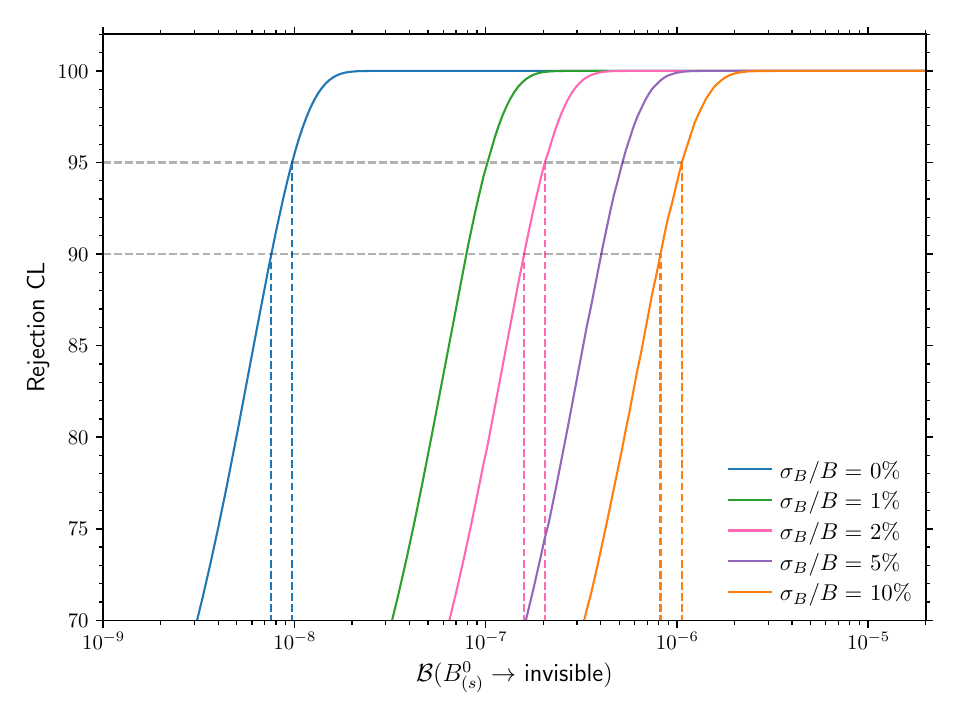}
    \caption{The impact of different systematic uncertainties on the anticipated constraints FCC-ee could set on \binvisible decays at different branching fractions, assuming negligible signal systematic.}
    \label{fig:significance_withsigmaB}
\end{figure}

As seen in Fig.~\ref{fig:significance}, systematic uncertainties hold the potential to significantly impact the sensitivity of FCC-ee to \binvisible decays. Therefore the effect of varying the background expectation systematic uncertainty, $\sigma_B$, was investigated assuming a negligible signal systematic, $\sigma_S\approx0$. Figure~\ref{fig:significance_withsigmaB} demonstrates the effect of introducing these systematic uncertainties via Eq.~\eqref{eq:FOMinclerr} to give an expected worst-case sensitivity. We note that to see an order of magnitude improvement on the branching fraction limits projected by \belletwo using the full $50\invab$ $\Upsilon(4S)$ dataset \cite{Belle-II:2018jsg}, $\sigma_B/B$ of approximately $2\%$ or less would be required.

\section{Conclusion}
\label{sec:conclusion}

We carry out a performance study on the measurement of \binvisible decays at FCC-ee running at the $Z$ pole.
Our studies demonstrate that FCC-ee offers an unparalleled and probably unique opportunity to search for these rare, experimentally difficult, yet highly new-physics sensitive decays.
We determine that branching fractions above $7.6\times10^{-9}$ ($9.7\times10^{-9}$) would be excluded at 90\% (95\%) confidence level, and branching fractions above $3.0\times10^{-8}$ would be within discovery reach, at FCC-ee if it produces $6\times 10^{12}$ $Z$ bosons.
These null searches for beyond SM physics offer a powerful and complementary probe to direct measurements of branching fractions and angular distributions in $\bquark\to\squark\neu\neub$ transitions, such as those reported in Ref.~\cite{Amhis:2023mpj}.
Should deviations from SM expectations be confirmed in $\bquark\to\squark\neu\neub$ decays, as potentially indicated by the recent \belletwo measurement~\cite{Belle-II:2023esi}, then measurements and improved limits on fully invisible \BdorBs decays will play an essential role in constraining the underlying NP scenarios. Furthermore, a broad class of well-motivated models, particularly those involving portals to dark sectors or long-lived neutral particles, can lead to sizeable enhancements in the \binvisible branching fraction without inducing significant modifications to SM $\bquark\to\squark\neu\neub$ transition rates.

\section*{Acknowledgments}

We thank the FCC-ee Physics Performance Group for the fruitful discussions and
helpful feedback on the analysis procedure and manuscript.
We also thank our colleagues in the Cambridge LHCb and Theory groups for the fun discussions and helpful advice. 
M.K, R.M and E.W are supported by UK Research and Innovation under grant \texttt{\#EP/X014746/2}.
R.M was supported by Clare College, Cambridge.
A.R.W. is supported by UK Research and Innovation under grant \texttt{\#MR/Y01166X/1}.

\bibliographystyle{LHCb}
\bibliography{references}

\end{document}